\documentclass[vecphys]{svmult}
\authorrunning{S. Phang, et al.}
\titlerunning{$\mathcal{PT}$-symmetric Grating Structures in Photonics}

\usepackage{makeidx}         

\usepackage{graphicx}\usepackage{epstopdf} 
\usepackage{overpic} 
\usepackage{rotating}

\usepackage{amsmath,esint}\usepackage{amssymb} \usepackage{mathrsfs} \usepackage{mathtools}  
\usepackage{bm} 

\usepackage{multicol}        
\usepackage{cite}            
\usepackage[bottom]{footmisc}

\usepackage{xcolor,calc, blindtext} 

\usepackage{booktabs}  

\makeindex             

\newcommand{\e}{{\rm e}}

\newcommand{\PT}{{$\mathcal{PT}$}}

\newcommand{\gainsat}{{$\mathbb{S}$ }} 
 
\newcommand{\posx}{{\bf x}} 
\newcommand{\momentum}{{\mathbf{\hat{p}}}}

\begin{document}
\setcounter{chapter}{5}

\title{Theory and Numerical Modelling of Parity-Time Symmetric Structures in Photonics:  Introduction and Grating Structures in One Dimension}
\author{S Phang\textsuperscript{1,2,*}, T M Benson\textsuperscript{2} \and H. Susanto\textsuperscript{3} \and S. C. Creagh\textsuperscript{1} \and G. Gradoni\textsuperscript{1,2} \and P. D. Sewell\textsuperscript{2} \and A. Vukovic\textsuperscript{2}}
\institute{
	Wave Modelling Research Group - School of Mathematical Sciences, University of Nottingham, United Kingdom
	\and George Green Institute for Electromagnetics Research, University of Nottingham, United Kingdom 
	\and Department of Mathematical Sciences, University of Essex, United Kingdom \\ 
	\textsuperscript{*} Corresponding author - \texttt{sendy.phang@nottingham.ac.uk}
}


\maketitle

\begin{abstract}
A class of structure based on \PT-symmetric Bragg gratings in the presence of both gain and loss is studied. The basic concepts and properties of parity and time reversal in one-dimensional structures that possess \textit{idealised} material properties are given.  The impact of \textit{realistic} material properties on the behaviour of these devices is then investigated. Further extension to include material non-linearity is used to study an innovative all-optical memory device.
\end{abstract}

\noindent

\section[Introduction]{Introduction of Parity and Time-Reversal (\PT) Symmetry}
\label{sec:intro}
\begin{figure}[t]
	\begin{overpic}[width=0.95\textwidth,tics=5]{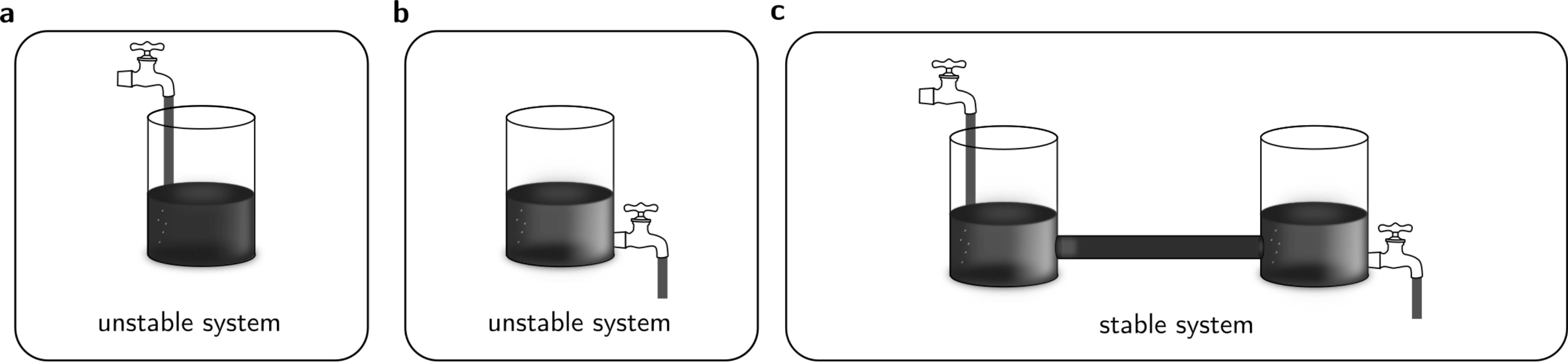}		
	\end{overpic}
	\centering
	\caption[Illustration of source-drain system]{ {Illustration of source-drain system. An isolated system with (a) source, (b) drain and (c) the coupled source-drain system.} }
	\label{fig:01_ilus}
\end{figure}

Studies of Parity-Time (\PT) symmetric \index{(\PT) symmetric} structures are motivated by a seminal paper by Bender and Boetcher \cite{Bender1998} in 1998. In the paper Bender and Boetcher introduced the concept of a \PT-symmetric Hamiltonian for Quantum Mechanical problems, in which it is established that a complex Quantum Mechanical Hamiltonian which satisfies a combined Parity and Time-reversal symmetry \textit{may} have a completely real spectrum, i.e. it is a stable system. Further studies \cite{Bender2002,wang2010} showed that a simple coupled source and drain \index{source and drain} problem with the following Hamiltonian 
\begin{align}
H = 
\begin{bmatrix}
-j\alpha &\kappa \\
\kappa &j\alpha 
\end{bmatrix}
\label{eq:pthami}
\end{align}
is in fact a subset of the large class of \PT-symmetric Hamiltonians. The eigenvalue problem matrix of the source and drain problem in Eq. (\ref{eq:pthami}) has $\alpha$ to represent the source ($+$) and drain ($-$) which are coupled by a coupling mechanism represented by $\kappa$. As such, in the simplest form, the concept of \PT-symmetry can be depicted as a source-drain system which is schematically illustrated in Fig. \ref{fig:01_ilus}. It can be seen from Fig. \ref{fig:01_ilus}(a) that a system with a source is unstable, in the same way as for a system with a drain, portrayed in Fig. \ref{fig:01_ilus}(b); the system with a source has a growing (unbounded) state while the system with a drain has a decaying state. It is, however, by coupling these systems together that a system with growing energy can be \textit{tamed} by a dissipating system which yields a stable system, as illustrated in Fig. \ref{fig:01_ilus}(c). 

Based on this simple concept, different physical systems have been employed to study the new class of \PT-symmetric physics, such as cold atom systems \cite{Dizdarevic2015,Gutohrlein2015,Single2014}, electronics \cite{Bagarello2013,Schindler2012,Schindler2011}, mechanical oscillators \cite{Bender2013}, acoustics\cite{Zhu2014,Fleury2015}, microwave electromagnetics \cite{Poli2015,Bittner2012} and optics-photonics \cite{Jones2012,Ramezani2010,Lin2011,Kulishov2013,Phang2013,Phang2014d,Phang2015,Huang2013,Rivolta2015,Regensburger2013,Longhi2011,Makris2008,Nolting1996,Ruschhaupt2005,Greenberg2005,sukhorukov2010,Nazari2011,Kuzmiak2010,Lupu2013,Benisty2011,Alaeian2014,Baum2015,Feng2014,Chang2014,Longhi2014,Peng2014,phang2015b,Peng2014d}. Within the area of optics-photonics the concept of \PT-symmetry has been considered in the context of gratings\cite{Jones2012,Ramezani2010,Lin2011,Kulishov2013,Phang2013,Phang2014d,Phang2015,Huang2013,Rivolta2015}, lattices \cite{Regensburger2013,Longhi2011,Makris2008}, waveguides \cite{Nolting1996,Ruschhaupt2005,Greenberg2005,sukhorukov2010,Nazari2011,Kuzmiak2010,Lupu2013}, plasmonics \cite{Benisty2011,Alaeian2014,Baum2015,Lupu2013} and resonant cavities \cite{Feng2014,Chang2014,Longhi2014,Peng2014,phang2015b,Peng2014d,Phang2015d}.      

This chapter will introduce the concept of Parity and Time structures in photonics and their Quantum Mechanics equivalences, and then summarise recent research studies of \PT-symmetric photonics with emphasis mainly on a \PT-symmetric Bragg grating structure. The following section presents a study of a \PT-symmetric Bragg grating structure, in which unique scattering phenomena, such as unidirectional invisibility \index{unidirectional invisibility} and Coherent-Perfect-Absorber and Lasing (CPAL) \index{simultaneous coherent-perfect-absorber-lasing} operations, have been discovered. This is followed by an investigation of the impact of \textit{realistic} gain/loss material on the features and properties of a \PT-Bragg grating. ``Realistic", here, means that we consider a simple three-level energy system with an homogeneously broadened dispersion profile, which is typically used to illustrate erbium-doped amplifier based material. A dispersive and saturable gain model, implemented within a time-domain Transmission-Line Modelling \index{Transmission-Line Modelling} (TLM) method is then introduced in detail. This enables the impact of realistic material properties on the behaviour of these device to be investigated. The TLM model is then further extended to include material non-linearity and used to study the behaviour of non-linear \PT-Bragg gratings as innovative all-optical memory device.   

Chapter 7 will  present a summary of recent studies in \index{\PT coupled microresonators} \PT coupled microresonators, such as the concept of lowering laser threshold by increasing loss. The emphasis of both of these book chapters is to model such structures (\PT-symmetric Bragg gratings \index{\PT-symmetric Bragg grating} and coupled microresonators) in the context of a realistic gain/loss material model which is non-linear and dispersive. Each chapter will presents conclusions and future perspectives for the development of \PT-symmetric photonics.

\section{Parity-Time (\PT) Symmetric Scatterers in 1-D}\label{sec:1dscatt}
	This section reviews the concept of \PT symmetry within the context of a Quantum Mechanical (QM) system. It is shown that a QM system with a \PT-symmetric Hamiltonian has a complex conjugated energy potential. By exploiting the isomorphism between the \index{Schr\"{o}dinger equation} Schr\"{o}dinger equation and the Helmholtz equation for a scattering system, an analogous \PT-symmetric photonic system is constructed by a judicious choice of complex dielectric parameters.
	\subsection{Parity and Time-Reversal ($\mathcal{PT}$) Symmetry} \label{sec::pt_schrodinger}
	In order to understand the concept of Parity and Time (\PT) symmetric structures in photonics, it is only natural to review some fundamental theorems and postulates in Quantum Mechanics (QM) in which the \PT-symmetric problem was firstly defined. In QM, it is well-known that the behaviour of a particle is described by the so-called Schr\"{o}dinger equation, the time-independent form of which is given by\cite{Zettili2009,Yariv1989},
	\begin{align}
	\hat{H} \psi = E \psi
	\label{eq:scroeq}
	\end{align}   
	where $\psi$ denotes the scalar time-independent wavefunction which is a function of position, i.e. $\psi(x,y,z)$ in the Cartesian coordinate system, $E$ refers to the eigenstates of the problem Eq. (\ref{eq:scroeq}) and $\hat{H}$ denotes the \index{Hamiltonian operator} Hamiltonian operator and has important roles as summarised below\cite{bender13a,bender13b,Bender1998}: 
	\begin{enumerate}
		\item To determine the energy eigenstates $E$ which essentially are the solutions of Eq. (\ref{eq:scroeq}). It implies that the energy eigenstates $E$ are the result of the action described by $\hat{H}$ applied on the state vector $\psi$. Moreover, considering that $E$ is a physically measurable quantity, it is essential for $E$ to be real.  \newline
		\item Within the context of the time-domain Schr\"{o}dinger equation, 
		\begin{align}
		\hat{H} \psi(x,y,z;t) = \text{i} \hslash \frac{\partial}{\partial t} \psi(x,y,z;t)
		\label{eq:timeschro}  
		\end{align}
		the Hamiltonian has a role to describe the time evolution of the state vector $\psi$ which is the time-domain solution of Eq. (\ref{eq:timeschro}). It is emphasised that the complex number notation $(\text{i}=-j)$ is used as is customary in Quantum Mechanics textbooks\cite{Zettili2009,Yariv1989}.     \newline
		\item The Hamiltonian incorporates symmetry properties into the theory. In QM the Hamiltonian may exhibit continuous symmetries, such as time and spatial translation, and discrete symmetries, such as parity inversion and time-reversal invariance \cite{bender13a,bender13b,Bender1998}. For example if the Hamiltonian commutes with the parity inversion symmetry, the Hamiltonian is said to be parity inversion invariant.     
	\end{enumerate}
	
	The Hamiltonian $\hat{H}$ is expressed in terms of the position $\posx$ and momentum $\momentum$ operator as, 
	\begin{align}
	\hat{H} = \momentum^2  + V(\posx)    
	\end{align}
	where, $\momentum$ and $V(\posx)$ denote the linear momentum operator and potential energy function of a particle, respectively. The linear momentum operator is imaginary and anti-symmetric, defined as $\momentum= -\text{i}\nabla$. It follows that $\momentum^2=-\nabla^2$ is real and symmetric (Hermitian) and therefore that \textit{if} the potential function $V({\posx})$ is a real function in space, it can be guaranteed that all the energy states $E$ are also real with the Hamiltonian $\hat{H}$ satisfying, 
	\begin{align}
	\hat{H} = \hat{H}^\dagger
	\label{eq:hermit}
	\end{align} 
	where $\dagger$ denotes a \index{Hermitian adjoint operation} Hermitian adjoint operation which in matrix form denotes a combined transpose and complex conjugation operation
	
	As suggested by Bender and Boettcher \cite{Bender1998}, although the Hermitian condition Eq. (\ref{eq:hermit}) is \textit{sufficient} to ensure all possible energy states to be completely real, it is not \textit{necessary}. In \cite{bender13a,bender13b,Bender1998}, it is further shown that a weaker symmetry than Hermiticity Eq. (\ref{eq:hermit}) may lead to real eigenvalues $E$, and this weaker symmetry is denoted as a \textit{Parity} ($\mathcal{P}$) and \textit{Time} ($\mathcal{T}$) symmetric Hamiltonian. As such the Hamiltonian $\hat{H}$ is invariant under the $\mathcal{PT}$ transformation,
	\begin{align}
	\mathcal{PT}\hat{H}\mathcal{PT} = \hat{H}
	\end{align}   
	where the parity operator $\mathcal{P}$ is defined as a linear operator which inverts space and momentum, and the time-reversal operator $\mathcal{T}$ is an operator which reverses time, i.e. $t\rightarrow-t$. The transformations performed by the parity and time-reversal operators are defined as\cite{bender13a,bender13b,Bender1998,Mostafazadeh2012,Ge2012,chong2011}, 
	\begin{align}
	\mathcal{P} &: {\posx} \rightarrow -{\posx} \quad ; \quad \momentum \rightarrow -\momentum \\
	\mathcal{T} &: j\rightarrow-j \quad ; \quad {\posx} \rightarrow {\posx} \quad ; \quad \momentum \rightarrow -\momentum
	\end{align}
	As such, it can be shown that a $\mathcal{PT}$-symmetric Hamiltonian in Quantum Mechanics is achieved when the potential function satisfies\cite{bender13a,bender13b,Bender1998,Mostafazadeh2012,Ge2012,chong2011},
	\begin{align}
	\mathcal{PT}V({\posx})\mathcal{PT} =V^*({-\posx}) = V({\posx}) 
	\label{eq:pthamil}
	\end{align}
	where $*$ denotes the conjugation operation. The \index{\PT-symmetric condition} \PT-symmetric condition Eq. (\ref{eq:pthamil}) implies that the energy potential $V(\posx)$ is a complex function where the real part is an even function and the imaginary part is an odd function in space.

	\subsection[$\mathcal{PT}$-Symmetric Photonics]{Photonics System Analogue of Quantum Mechanics $\mathcal{PT}$-Symmetric Hamiltonian} \label{sec::pt_hemlholtz}
	In contrast to the Schr\"{o}dinger equation in Quantum Mechanics, in optics-photonics the dynamics of an electromagnetic field are defined by the \index{Helmholtz equation} Helmholtz equation  which for the electric field, the Helmholtz equation is given as,
	\begin{align}
	\nabla^2\bm{\mathcal{E}} +\frac{\omega^2}{c_0^2} \bar{\varepsilon}(\posx) \bm{\mathcal{E}} = 0
	\label{eq:helmEeq}
	\end{align} 
	where $\bar{\varepsilon}(\posx)$ is the relative permittivity of the material and is a function of space $\posx$, such that it can be expressed in the form of, 
	\begin{align}
	\bar{\varepsilon}(\posx) = \bar{\varepsilon}_b + \Delta\bar{\varepsilon}(\posx)
	\label{eq:refidx}
	\end{align}
	In Eq. (\ref{eq:refidx}), $\bar{\varepsilon}_b$ denotes the homogeneous background material relative permittivity on which the spatial modulation $\Delta\varepsilon(\posx)$ occurs. By substituting the permittivity profile function Eq. (\ref{eq:refidx}) to Eq. (\ref{eq:helmEeq}), the Helmholtz equation can also be formulated as\cite{Longhi2011}, 
	\begin{align}
	\left\lbrace \nabla^2 +\frac{\omega^2}{c_0^2} \Delta\bar{\varepsilon}(\posx) \right\rbrace \bm{\mathcal{E}} =
	-\frac{\omega^2}{c_0^2} \bar{\varepsilon}_b \bm{\mathcal{E}}
	\label{eq:helmeqmod}
	\end{align}   
	
	\begin{table}[b]
		\centering
		\caption{{Comparison of the Helmholtz and Schr\"{o}dinger equations}.}
		\begin{tabular}{@{}lcc@{}} 
			\toprule
			& Quantum Mechanics        & Electromagnetics  \\ 
			\midrule
			Field              & $\Psi(\posx,t)=\psi(\posx)\e^{jEt/\hslash}$  & $\bm{E}(\posx;t) = \mbox{Re}\left[ \bm{\mathcal{E}}(\posx) \e^{j\omega t} \right]$ \\
			Eigenvalue problem & $\hat{H} \psi = E \psi $ & $\hat{\Theta} \bm{\mathcal{E}} = -(\frac{\omega}{c_0})^2 \varepsilon_b\bm{\mathcal{E}} $ \\
			Hamiltonian        & $\hat{H} = \momentum^2  + V(\posx)   $  & $\hat{\Theta}=\nabla^2 +(\frac{\omega}{c_0})^2 \Delta\bar{\varepsilon}(\posx)$\\		\bottomrule
		\end{tabular}
		\label{tab:emscheq}
	\end{table}
	
	By comparing Eq. (\ref{eq:helmeqmod}) and Eq. (\ref{eq:scroeq}), it can be seen that the time-harmonic Helmholtz equation of wave dynamics, albeit multicomponent, is isomorphic with the time-independent Schr\"{o}dinger Eq. (\ref{eq:scroeq}). The comparison of the Schr\"{o}dinger and Helmholtz equations is summarised in Table \ref{tab:emscheq}. Based on this analogy, it can be shown that \PT-symmetric photonic structure has a dielectric profile that satisfies,  
	\begin{align}
	\bar{\varepsilon}(\posx) =\bar{\varepsilon}^*(-\posx) \quad \text{or} \quad
	n(\posx) = n^*(-\posx)
	\label{eq:ptrefidx}
	\end{align} 
	so that the real part of permittivity (or refractive index) is an even function and the imaginary part of the permittivity (or refractive index) is an odd function of space, 
	\begin{align}
	\varepsilon'(-\posx) &= \varepsilon'(\posx) \\
	\varepsilon''(-\posx) &= -\varepsilon''(\posx)
	\label{eq:ptsymmatpho}
	\end{align}
	As such Eq. (\ref{eq:ptsymmatpho}) implies that a \PT-symmetric structure in photonics requires the presence of both \index{gain and loss} gain and loss in the system.  
	
	\subsection{Generalised Conservation Relations} \label{eq:genconsrel}
\begin{figure}[t]
		\centering
		\begin{overpic}[width=0.6\textwidth,tics=5] {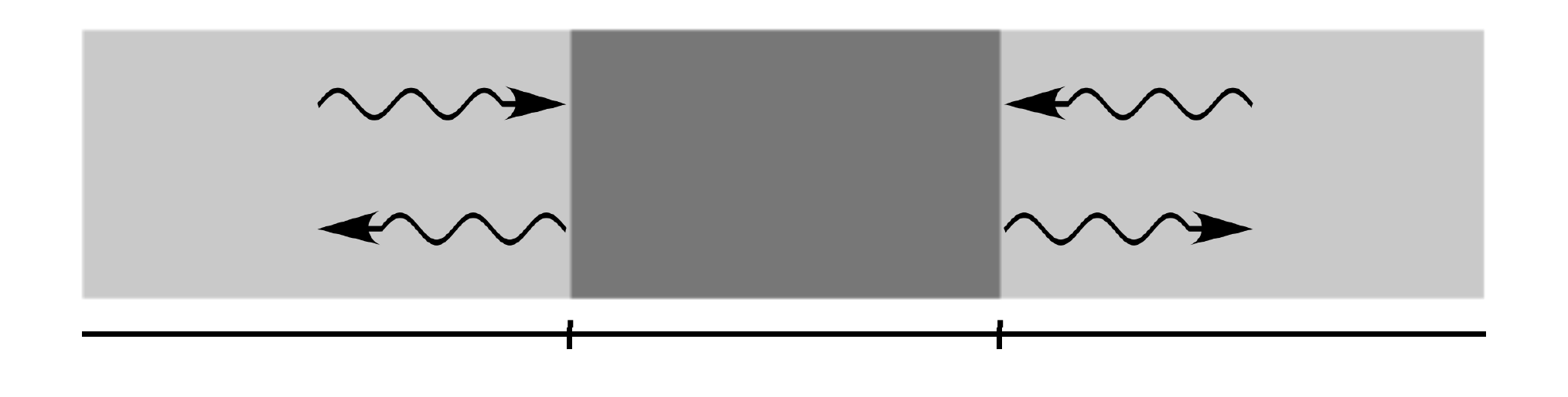}
			\put(45,13){\large $n(x)$} \put(7,15){$n_b$} \put(89,15){$n_b$}
			\put(15,19){$a_1$} \put(15.5,10){$b_1$}
			\put(82,19){$a_2$} \put(82,10){$b_2$}
			\put(31.5,-1.){$-\frac{L}{2}$} \put(63.2,-1.){$\frac{L}{2}$}
		\end{overpic}
		\caption[Schematic illustration of a one-dimensional scattering system.]{ {Schematic illustration of a one-dimensional scattering system.} }
		\label{fig:04_pt1dscatt}
	\end{figure}
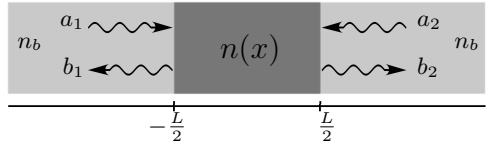

	For definiteness, consider a 1D \PT-symmetric structure schematically illustrated in Fig. \ref{fig:04_pt1dscatt}. The structure has a length $L$ with a refractive index profile satisfying the \PT-symmetric condition in the longitudinal direction, i.e. $n^*(x)=n(-x)$, and is embedded in a lossless background material of refractive index $n_b$. In Fig. \ref{fig:04_pt1dscatt}, the incoming $a$ and outgoing $b$ wave amplitudes are denoted for both the left and right sides. The longitudinal-components of the electric field on each port can be expanded as, 
	\begin{align}
	\mathcal{E}_T(x) =
	\begin{cases}
	a_1 \e^{j\beta x} + b_1 \e^{-j\beta x}   \quad \quad \text{for : } x<-\frac{L}{2} \\
	a_2 \e^{j\beta x} + b_2 \e^{-j\beta x}   \quad \quad \text{for : }  x>\frac{L}{2}
	\end{cases} 
	\end{align}
	
	As such the \index{wave-scattering} wave-scattering can be modelled by the \index{$\mathbf{S}$-matrix} $\mathbf{S}$-matrix as, 
	\begin{align}
	\begin{bmatrix}
	b_1 \\
	b_2
	\end{bmatrix} = 
	\mathbf{S}
	\begin{bmatrix}
	a_1 \\
	a_2
	\end{bmatrix} \quad\text{where, } \quad
	\mathbf{S} = 
	\begin{bmatrix}
	r_L  & t_R \\
	t_L  & r_R 
	\end{bmatrix}
	\label{eq:scatmat0}
	\end{align}
	and the quantities in the $\mathbf{S}$-matrix are defined as, 
	\begin{align*}
	\begin{split}
	t_L \quad &\text{:} \quad \text{transmission coefficient for left incidence}, \\
	t_R \quad &\text{:} \quad \text{transmission coefficient for right incidence}, \\
	r_L\quad &\text{:} \quad \text{reflection coefficient for left incidence}, \\
	r_R\quad &\text{:} \quad \text{reflection coefficient for right incidence}. 
	\end{split}
	\end{align*}
	
\begin{figure}[t]
		\centering
		\begin{overpic}[width=0.6\textwidth,tics=5] {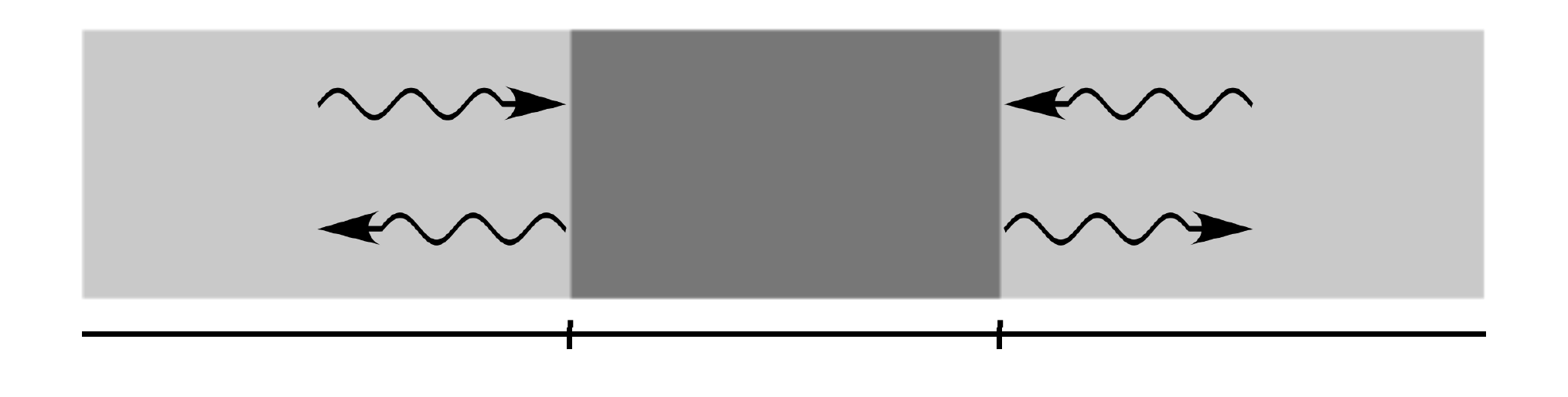}
			\put(45,14){\large $n(x)$} \put(8,15){$n_b$} \put(88,15){$n_b$}
			\put(15,18){$b^*_2$} \put(15.5,9){$a^*_2$}
			\put(82,18){$b^*_1$} \put(82,9){$a^*_1$}
			\put(31.5,-1.){$-\frac{L}{2}$} \put(63.2,-1.){$\frac{L}{2}$}
		\end{overpic}
		\caption[\PT-transformed scattering system.]{$ {\mathcal{PT}}$ {-transformed scattering system. The original system before \PT-symmetry transformation is illustrated in Fig. \ref{fig:04_pt1dscatt}.}}
		\label{fig:04_pt1dscatttransformed}
	\end{figure}

	If linear and non-magnetic materials are considered, \index{Lorentz reciprocity} Lorentz reciprocity holds, i.e. $\mathbf{S}=\mathbf{S}^T$; the $\mathbf{S}$-matrix can be simplified based on the reciprocality of left and right transmission coefficients, $t_L=t_R\equiv t$, as, 
	\begin{align}
	\mathbf{S} = 
	\begin{bmatrix}
	r_L  & t \\
	t     & r_R 
	\end{bmatrix}
	\label{eq:scatmat2}
	\end{align}
	From Eq. (\ref{eq:scatmat0}) and Eq. (\ref{eq:scatmat2}), the transfer matrix \index{$\mathbf{M}$-matrix} $\mathbf{M}$-matrix associated with Fig. \ref{fig:04_pt1dscatt}, which relates the left and right wave amplitudes, could be constructed as:   
	\begin{align}
	\begin{bmatrix}
	a_1 \\
	b_1
	\end{bmatrix} =
	\mathbf{M}
	\begin{bmatrix}
	b_2 \\
	a_2
	\end{bmatrix}
	\quad \text{where, } \quad
	\mathbf{M} \equiv
	\begin{bmatrix}
	\frac{1}{t} & -\frac{r_R}{t} \\
	\frac{r_L}{t} & t-\frac{r_Lr_R}{t}
	\end{bmatrix}
	\label{eq:Mmatrixori}
	\end{align}
	Moreover since the structure is \PT-symmetric, the \PT-transformed solutions should also be solutions of the Helmholtz equation. As such, Fig. \ref{fig:04_pt1dscatttransformed} depicts the $\mathcal{PT}$ transformed solution of the original problem in Fig. \ref{fig:04_pt1dscatt}. The \PT-transformed solutions are expressed as, 
	\begin{align}
	\mathcal{PT}\left\lbrace \mathcal{E}_T(x) \right\rbrace  = 
	\begin{cases}
	b^*_2 e^{j\beta x} + a^*_2 e^{-j\beta x}   \quad \quad \text{for : } x<-\frac{L}{2} \\
	b^*_1 e^{j\beta x} + a^*_1 e^{-j\beta x}   \quad \quad \text{for : }  x>\frac{L}{2}
	\end{cases}
	\end{align} 
	
	The corresponding $\mathbf{M}$-matrix formulation is now given by,
	\begin{align}
	\begin{bmatrix}
	b^*_2 \\
	a^*_2
	\end{bmatrix} =
	\mathbf{M}^{(\mathcal{PT})}
	\begin{bmatrix}
	a^*_1 \\
	b^*_1
	\end{bmatrix}
	\label{eq:Mmattransformed}
	\end{align}
	By a direct comparison of Eq. (\ref{eq:Mmatrixori}) and Eq. (\ref{eq:Mmattransformed}), the corresponding \PT-transformed matrix $\mathbf{M}$ is\cite{Ge2012,chong2011,longhi2010a,ISI:000266685400002}, 
	\begin{align}
	\mathbf{M} \xrightarrow{\mathcal{PT}} \mathbf{M}^{(\mathcal{PT})} \equiv
	\mathbf{M}^{-1*} 
	\end{align}
	Since the structure is \PT-symmetric invariant, it can be deduced that, 
	\begin{align}
	\mathbf{M} = \mathbf{M}^{-1*}
	\label{eq:ptMsymm}
	\end{align}
	Exploiting the fact that $\det(\mathbf{M})=1$, it can be shown that,
	\begin{align*}
	M_{11} = M^*_{22} \quad \text{ and } \quad \text{Re}[M_{12}] = \text{Re}[M_{21}] \equiv 0
	\end{align*}
	Using these relations, the $\mathbf{M}$-matrix can be parameterised as, 
	\begin{align}
	\mathbf{M} =
	\begin{bmatrix}
	A & -jB \\
	jC & A^*
	\end{bmatrix} \quad \text{ where } \quad \left\lbrace B,C \right\rbrace  \in \mathbb{R}
	\label{eq:ptMmatrix}
	\end{align}
	Here, each entry of the $\mathbf{M}$-matrix is defined as, 
	\begin{subequations}
		\begin{align}
		A &= \frac{1}{t} = t^*-\frac{r^*_Lr^*_R}{t^*}  \\
		B &= -j\frac{r_R}{t} = j\frac{r^*_R}{t^*}  \\
		C &= -j\frac{r_L}{t} = j\frac{r^*_L}{t^*} 
		\end{align}
		\label{eq:Mmatparameter}
	\end{subequations}
	By further exploitation of $\det(\mathbf{M})=1$ on Eq. (\ref{eq:ptMmatrix}), the \textit{generalised conservation relation} \index{generalised conservation relation} is formulated as\cite{Ge2012,chong2011,longhi2010a,ISI:000266685400002}, 
	\begin{align}
	1-|t|^2 = r_Lr^*_R = r^*_Lr_R 
	\label{eq:genconsrelation}
	\end{align}
	From Eq. (\ref{eq:genconsrelation}) the generalised conservation relation can also be expressed in terms of the transmittance $T=|t|^2$ and reflectance $R_{L,R} = |r_{L,R}|^2$ coefficients as\cite{Ge2012,chong2011,longhi2010a,ISI:000266685400002}, 
	\begin{align}
	|1-T| = \sqrt{R_LR_R}  
	\label{eq:genconsvrel}
	\end{align}  
	Generally, the conservation relation Eq. (\ref{eq:genconsvrel}) implies that one the following cases may occur:
	\begin{enumerate}
		\item For the case of $T<1$, Eq. (\ref{eq:genconsvrel}) reduces to $T+\sqrt{R_LR_R}=1$. It can be seen that $\sqrt{R_LR_R}$ replaces the conventional $R$ in the case of an orthogonal system. It follows that when $T<1$ the scattering of a single incident wave from one side of the structure yields to a loss of power flux\cite{Ge2012}. This operation is referred to throughout as \textit{sub-unitary transmission} \index{sub-unitary transmission} operation. Moreover, from Eq. (\ref{eq:genconsrelation}) one can evaluate the phase relation between the left and right reflected light. Consider that the reflected signal from left side is of the form of $r_{L}=|r_{L}|\e^{j\phi_{L}}$ and from the right side is of the form  $r_{R}=|r_{R}|\e^{j\phi_{R}}$. From Eq. (\ref{eq:genconsrelation}) it can be found that the phase for the left and right reflected signal is related by $\phi_L=\phi_R$.
		
		\item For the case of $T>1$, Eq. (\ref{eq:genconsvrel}) reduces to $T-\sqrt{R_LR_R}=1$. In this case, a single incident beam yields to a \textit{super-unitary transmission} \index{super-unitary transmission} with the phase relation between left and right reflected waves as $\phi_L-\phi_R=\pi$.
		
		\item For the case of $T=1$, Eq. (\ref{eq:genconsvrel}) reduces to $\sqrt{R_LR_R}=0$ which implies that the product of the left and right reflectances must be zero. Such an operation is typically accomplished by having no reflection from one side of the structure. This particular operation case is referred to as \textit{unidirectionally invisible} \index{unidirectional invisibility} operation. 
	\end{enumerate}
	
	\subsection{Phases in a $\mathcal{PT}$ Scattering System}
	\label{sec:pttransition}
	The relationship between the incoming and outgoing waves in an optical network is well-described by using the scattering matrix $\mathbf{S}$. This section will focus on investigating the spectral properties of the $\mathbf{S}$-matrix associated with the $\mathcal{PT}$-symmetric scattering system depicted in Fig. \ref{fig:04_pt1dscatt}. 
	
	It is well-understood that for any linear passive structure, i.e. no gain and loss, the $\mathbf{S}$-matrix is unitary \cite{Benson1991,Ramo199,Pozar2011,Haus1983,Collin1991}, 
	\begin{align}
	\mathbf{S}^\dagger = \mathbf{S}^{-1} 
	\label{eq:unitaryS}
	\end{align}
	where $\dagger$ denotes the transpose and conjugation operation. The unitary relation \index{unitary relation} Eq. (\ref{eq:unitaryS}) puts a strict condition that the eigenvalues $s_n$ of the $\mathbf{S}$-matrix have to be unimodular, i.e.
	\begin{align*}
	|s_n| = 1 
	\end{align*}
	Hence, for a passive structure, power is conserved with no net amplification or dissipation. 
	
	However, in the case when a gain or/and lossy element is present, as is the case in \PT-symmetric structures, the spectral behaviour of the $\mathbf{S}$-matrix is non-trivial. It will be shown shortly that the \PT-symmetric structure may undergo a \textit{phase transition} \index{phase transition} from a power conserving operation, with no net-amplification or dissipation, to a non-conserving system. Hence, consider the scattering matrix formulation associated with Fig. \ref{fig:04_pt1dscatt} expressed as, 
	\begin{align}
	\begin{bmatrix}
	b_1 \\
	b_2
	\end{bmatrix} = 
	\mathbf{S}
	\begin{bmatrix}
	a_1 \\
	a_2
	\end{bmatrix} \quad\text{where, } \quad
	\mathbf{S} = 
	\begin{bmatrix}
	r_L  & t \\
	t  & r_R 
	\end{bmatrix}
	\label{eq:scatmat}
	\end{align}
	where the ingoing and outgoing field amplitudes ($a$ and $b$) for each individual port can also be expressed compactly as, 
	\begin{align}
	\mathcal{E}_T(x) &= \sum_{n=1}^{2} [a_n \e^{-j\beta_n x} + b_n \e^{j\beta_n x}]  
	\end{align}
	Following similar reasoning to that employed in Section \ref{eq:genconsrel}, a \PT-symmetric scattering system should support the \PT-transformed solution on each port, which is 
	\begin{align}
	\mathcal{PT}\left\lbrace \mathcal{E}_T(x)\right\rbrace  =  \sum_{n=1}^{2} [(\mathcal{PT} a_n) \e^{j\beta_n x} + (\mathcal{PT}b_n) \e^{-j\beta_n x}]. 
	\end{align}
	As such the following scattering formulation is also valid, 
	\begin{align}
	\mathcal{PT}
	\begin{bmatrix}
	a_1 \\
	a_2
	\end{bmatrix} = 
	\mathbf{S}
	\mathcal{PT}
	\begin{bmatrix}
	b_1 \\
	b_2
	\end{bmatrix} 
	\label{eq:pttransS}
	\end{align}
	where in the matrix formulation the operators $\mathcal{P}$ and $\mathcal{T}$ are defined as
	\begin{align}
	\mathcal{P} = 
	\begin{bmatrix}
	0  & 1 \\
	1  & 0 
	\end{bmatrix} \quad \text{and} \quad \mathcal{T} = \mathcal{K}
	\end{align}
	where $\mathcal{K}$ is the conjugation operation \index{conjugation operation}. By comparing Eq. (\ref{eq:scatmat}) and Eq. (\ref{eq:pttransS}), it can be found that the scattering matrix $\mathbf{S}$ obeys the following \PT-symmetric transformation, 
	\begin{align}
	\mathcal{PT} \mathbf{S} \mathcal{PT} = \mathbf{S}^{-1}    
	\label{eq:ptSmattransf}
	\end{align} 
	For convenience, consider the parameterised $\mathbf{S}$-matrix using Eq. (\ref{eq:Mmatparameter}) as,
	\begin{align}
	\mathbf{S} = 
	\frac{1}{A}
	\begin{bmatrix}
	jC & 1 \\
	1 & jB
	\end{bmatrix}  \quad \text{where, } \quad \{B,C\} \in \mathbb{R} 
	\end{align} 
	By direct calculation, the eigenvalues $s_n$, with $n\in\{1,2\}$, can be expressed as \cite{Ge2012}, 
	\begin{align}
	s_1,s_2 = \frac{j}{2A} \left[ (B+C) \pm \sqrt{(B-C)^2-4}\right]  \quad \text{where, } \quad \{B,C\} \in \mathbb{R} 
	\end{align} 
	Since the parameters $B$ and $C$ are real numbers, it can be deduced that one of the following cases may happen, 
	\begin{enumerate}
		\item For the case of $(B-C)^2<4$, the eigenvalues are
		\begin{align}
		s_1,s_2 = \frac{j}{2A} \left[ (B+C) \pm j\sqrt{4-(B-C)^2}\right], \quad \{B,C\} \in \mathbb{R} 
		\label{eq:eigvalptsym}
		\end{align} 
		and the corresponding eigenvectors are, 
		\begin{align}
		\bm{\psi}_1,\bm{\psi}_2 = \begin{pmatrix}
		2+ j \left[ (C-B) \pm j\sqrt{4-(B-C)^2} \right]  \\
		2+ j \left[ (B-C) \pm j\sqrt{4-(B-C)^2} \right] 
		\end{pmatrix} \quad \text{for: } \quad s_1,s_2
		\end{align}
		From Eq. (\ref{eq:eigvalptsym}),  it can be found by direct calculation that the eigenvalues are unimodular, i.e. $|s_n|=1$. This implies that, for this case, power is conserved thus there is no net-amplification nor dissipation. Note that, in this case, the eigenvectors themselves are \PT-symmetric as the \PT-operation transforms the eigenvectors back to themselves,
		\begin{align}
		\bm{\psi}_1,\bm{\psi}_2  \xrightarrow{\mathcal{PT}}  \bm{\psi}_1,\bm{\psi}_2  
		\end{align}
		This particular operation case is referred to as the \textit{\PT-symmetric phase} \index{\PT-symmetric phase}. \newline
		\item For the case of $(B-C)^2>4$, the eigenvalues are
		\begin{align}
		s_1,s_2 = \frac{j}{2A} \left[ (B+C) \pm \sqrt{(B-C)^2-4}\right], \quad \{B,C\} \in \mathbb{R} 
		\end{align} 
		with the corresponding eigenvectors as, 
		\begin{align}
		\bm{\psi}_1,\bm{\psi}_2 = \begin{pmatrix}
		2+ j \left[ (C-B) \pm \sqrt{(B-C)^2-4} \right]  \\
		2+ j \left[ (B-C) \pm \sqrt{(B-C)^2-4} \right] 
		\end{pmatrix} \quad \text{for: } \quad s_1,s_2  
		\end{align} 
		and the transformed solution is 
		\begin{align*}
		\mathcal{PT} \left\lbrace \bm{\psi}_1,\bm{\psi}_2 \right\rbrace   = \begin{pmatrix}
		2 + j \left[ (C-B) \mp \sqrt{(B-C)^2-4} \right]  \\
		2 + j \left[ (B-C) \mp \sqrt{(B-C)^2-4} \right] 
		\end{pmatrix} \quad \text{for: } \quad s_1,s_2 
		\end{align*} 
		Hence, it can be seen that, in this case, the eigenvectors are not \PT-symmetric but the pair satisfies the \PT-transformation, by transforming to each other,  
		\begin{align}
		\bm{\psi}_1,\bm{\psi}_2  \xrightarrow{\mathcal{PT}}  \bm{\psi}_2,\bm{\psi}_1  
		\end{align}
		Exploiting the symmetry properties of the \PT-symmetric $\mathbf{S}$-matrix Eq. (\ref{eq:ptSmattransf}), it can be deduced that the pair of eigenvalues are reciprocally conjugate, i.e.
		\begin{align}
		s^*_{1,2}s_{2,1}  = 1 
		\label{eq:productunit}
		\end{align}
		which implies in  general that if $|s_1|>1$ then $|s_2|<1$. Operation in this case is denoted as the \textit{\PT-broken symmetry phase} \index{\PT-broken symmetry phase}. \newline 
		\item The case of $(B-C)^2=4$, is the case when both of the above cases are true. In this case, one can find that the eigenvalues are degenerate, as
		\begin{align}
		s_1=s_2 \equiv \frac{j}{2A}  (B+C) = \pm j\frac{|A|}{A}   \quad \text{where, } \quad \{B,C\} \in \mathbb{R}   
		\end{align}
		with the associated eigenvectors, 
		\begin{align}
		\bm{\psi}_1=\bm{\psi}_2 \equiv \begin{pmatrix}
		1 \pm j    \\
		1 \mp  j   
		\end{pmatrix} \quad \text{for: } \quad s_1,s_2   
		\end{align}
		This particular point is referred to as the \textit{\PT-breaking point} \index{\PT-breaking point}. 
	\end{enumerate} 
	
	It has been shown that the eigenvalues of the $\mathbf{S}$-matrix characterise the operational phase of a \PT-symmetric system. As such when the eigenvalues are unimodular, the system is in the \PT-symmetric phase whilst if the eigenvalues are reciprocal conjugate the system is in the \PT-broken symmetry phase. A simpler criterion can be obtained by evaluating the value of $(B-C)^2$ using the identities given in Eq. (\ref{eq:Mmatparameter}). The criterion of \PT-symmetry phases can be expressed in terms of the transmittance and reflectance coefficients as\cite{Ge2012},
	\begin{align}
	\frac{R_L +R_R}{2} -T 
	\begin{cases}
	< 1 , \quad \text{for : \textit{\PT-symmetric phase}} \\
	= 1 , \quad \text{for : \textit{\PT-symmetry breaking point}} \\ 
	> 1 , \quad \text{for : \textit{\PT-broken symmetric phase}} 
	\end{cases}
	\label{eq:ptcriterion}
	\end{align} 
	where, $T=|t|^2$ denotes the transmittance and $R_{L,R}=|r_{L,R}|^2$ denotes the reflectance coefficients for the incident wave arriving from the left or the right of the structure respectively.   
	
	\subsection{Simultaneous Coherent Perfect Absorber and Lasing}
	\label{sec:cpalsection}
	A simultaneous lasing and absorbing \index{simultaneous coherent-perfect-absorber-lasing} action in a \PT-symmetric scatterer was noted in \cite{ISI:000266685400002,Ge2012,chong2011,longhi2010a}. In order to understand the properties of this operation, first consider a laser system. In a laser system, the structure emits light even in the absence of an injected signal beam hence $a_1=a_2=0$ whilst $\{b_1,b_2\}\rightarrow\infty$. Imposing this solution upon the $\mathbf{M}$-matrix Eq. (\ref{eq:Mmatrixori}) means that the entry $M_{11}=0$. On the other hand, if a structure behaves as a coherent perfect absorber (CPA), there is no scattered wave $b_1=b_2=0$ whilst $\{a_1,a_2\}\ne0$, and this requires the entry $M_{22}=0$. In practice, the conditions $M_{11}=0$ and $M_{22}=0$ do not coincide at the same frequency, hence the system is either lasing or absorbing, but not both.  
	
	However, in a \PT-symmetric structure the condition Eq. (\ref{eq:ptMsymm}) does allow this to happen as $M_{11}=M_{22}=0$ can occur generically at the same frequency. Moreover, since $\det(\mathbf{M})=1$, it can be further shown that, 
	\begin{align}
	M_{11}=M_{22}=0 \quad \text{and} \quad M_{12},M_{21} \ne 0 
	\end{align}
	which implies that transmission and reflection from the both sides are such that, 
	\begin{align}
	|t|\rightarrow\infty \quad \text{and} \quad |r_L| , |r_R| \rightarrow \infty 
	\label{eq:cpaltr}
	\end{align}
	and their phase can be calculated from Eq. (\ref{eq:Mmatrixori}) as, 
	\begin{align}
	\phi_R+\phi_L = 2\phi_t
	\end{align}
	where $\phi_{L}$ and $\phi_{R}$ denote the phase for the reflected wave for the signal incident from left and right side for the structure respectively whilst $\phi_t$ is the phase of the transmitted wave.
	
	The scenario of Eq. (\ref{eq:cpaltr}) can only occur in the \PT-broken symmetry phase, i.e. operation case 2 described in Section \ref{sec:pttransition}. This implies that one of the eigenvalues of the $\mathbf{S}$-matrix is $|s_1|\rightarrow0$ whilst the other one is $|s_2|\rightarrow\infty$, noting that the product of them should remain unity Eq. (\ref{eq:productunit}). This physically means that the structure has both lasing and coherent perfect absorber(CPA) states simultaneously. Such operation is referred to as \textit{simultaneous coherent perfect absorber-lasing}(CPAL) \index{simultaneous coherent-perfect-absorber-lasing} operation. Since $|\det(\mathbf{S})|=|s_1s_2|$, one can also interpret that CPAL occurs when the poles and zeros of the $\mathbf{S}$-matrix coalesce in the real frequency axis.         
	
	\subsection[Bragg Grating with a \PT-Symmetric Refractive Index Modulation]{Bragg Grating with a \PT-Symmetric Refractive Index Modulation} 
	\label{sec:noncausalPTBG}
	
	In this section, let us consider a \PT-symmetric Bragg grating (PTBG) \index{\PT-symmetric Bragg grating}, i.e. a Bragg grating structure with a \PT-symmetric refractive index modulation profile. This section will focus on the study of the effect of the \PT-symmetric phase transition and the spectral singularity on the operation of the PTBG. Special attention will be given to different kinds of transmission, i.e. the sub-unitary, super-unitary and unitary transmissions which were described in Subsection \ref{eq:genconsrel}. 
	
	\begin{figure}[t]
		\begin{overpic}[width=0.8\textwidth,tics=5]{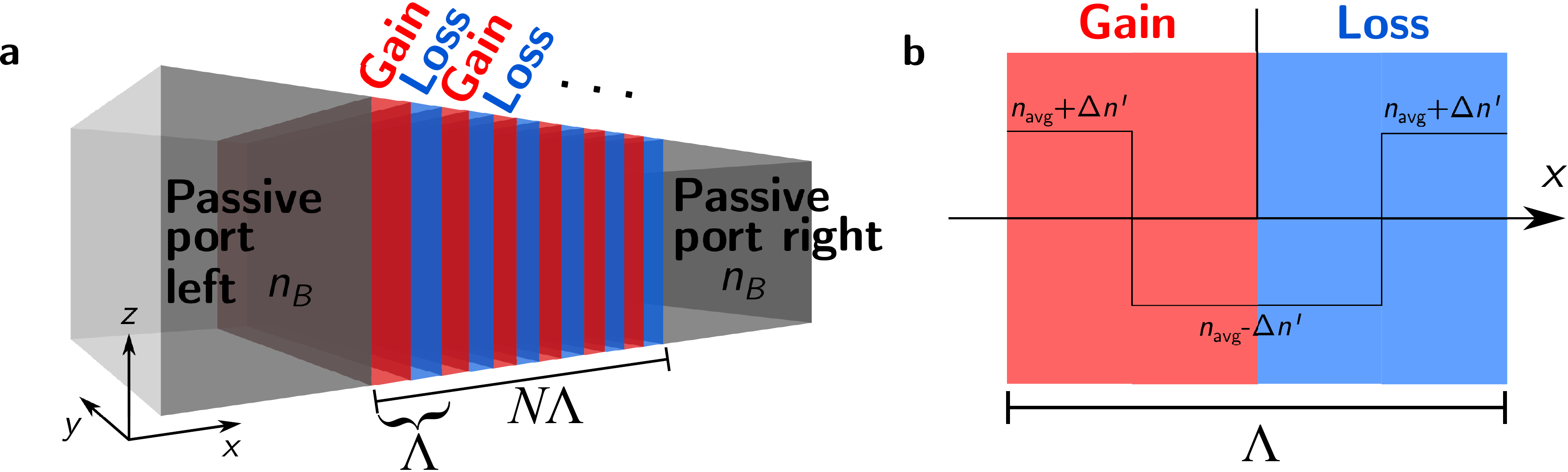}
		\end{overpic}
		\centering
		\caption[Schematic of a \PT-Bragg grating structure.]{ {Schematic of a \PT-Bragg grating structure. ({a}) Grating composed of $N$ unit cell in a background material $n_B$, ({b}) single unit cell of the grating with 2 slightly different refractive indices $n_\text{avg}+\Delta n'$ and $n_\text{avg}-\Delta n'$. Red coloured sections denote gain while the lossy sections are coloured blue.}}
		\label{fig:05_ptbgillus}
	\end{figure}
	
	The \PT-symmetric refractive index modulation requires that the real part of the refractive index is an even function of position and the imaginary part of the refractive index, which represents gain and loss, is an odd function of position. The \PT-symmetric Bragg grating (PTBG) considered has piecewise constant layers of refractive index $n=(n_\text{avg}\pm \Delta n') \pm jn''$, where $n_\text{avg}$ is the average refractive index, and $\Delta n'$ and $n''$ are the modulations of the real and imaginary parts of the refractive index respectively. The grating is surrounded by a background material of average refractive index $n_b=n_\text{avg}$ and has total length of $N\Lambda$, where $\Lambda$ is the length of one unit cell and $N$ is the number of unit cells. This is schematically illustrated in Fig. \ref{fig:05_ptbgillus}.  
	
\begin{figure}[]	
		\centering
		\begin{overpic}[width=0.8\textwidth,tics=5]{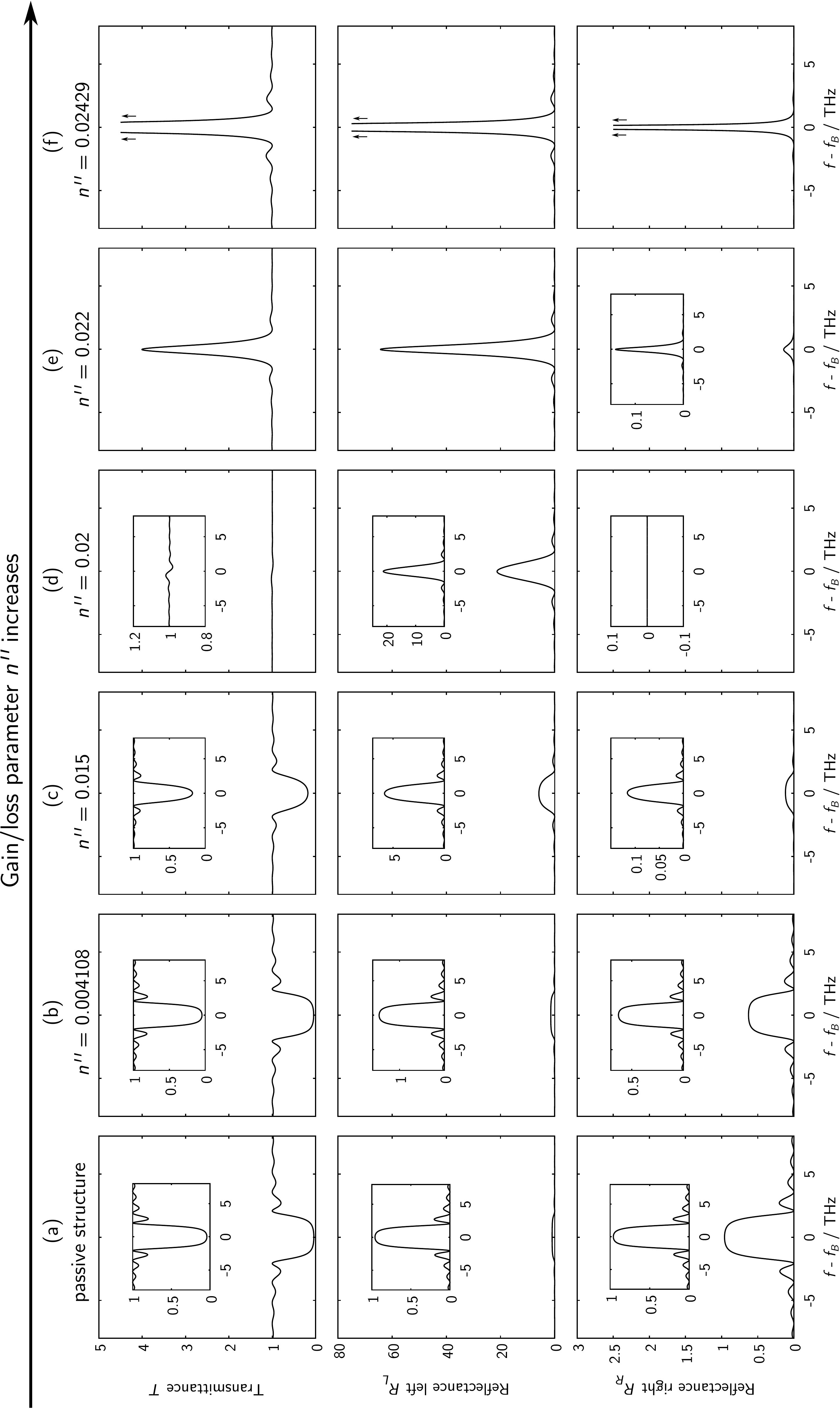}
		\end{overpic}
		\caption[Transmittance and reflectance spectra of a PTBG.]{ {Transmittance and reflectance spectra of a PTBG. The transmittance $T$, reflectance for the left and right incident waves ($R_L$ and $R_R$ respectively) are displayed in the top, middle and bottom panels respectively. Six different values of gain/loss parameter $n''=0$,  $0.0041$, $0.015$, $0.02$, $0.022$ and $0.02429$ are considered. The insets show amplified details of spectra.} }
		\label{fig:05_ptbg_t_r_evolve}
	\end{figure}

	For definiteness, consider a PTBG with a depth of real part modulation of $\Delta n'=0.02$ that is designed with a Bragg frequency $f_B=336.845$ THz ($\lambda_B=0.89$ $\mu$m) and that the number of periods of $N=200$. Moreover, the background material and the average refractive index of the structure are taken as $n_b=n_\text{avg}=3.5$, a value typical of a semiconductor material. The pitch length of a single unit cell of the grating is calculated as $\Lambda=\lambda_B/(2n_\text{avg})=0.127\text{ }\mu$m. The transmittance and reflectance for both left and right incident waves  are plotted in Fig. \ref{fig:05_ptbg_t_r_evolve} for an increasing value of gain/loss parameter, i.e. $n'' = 0$, $0.0041$, $0.015$, $0.02$, $0.022$ and $0.02429$. They are calculated by the analytical Transfer matrix \index{Transfer matrix method} (T-matrix) method, We are not describing the T-matrix method in this chapter; for detail on the method readers are referred to \cite{Collin1991,Pozar2011}.  
	
	The Lorentz reciprocity \index{Lorentz reciprocity} theorem states that the $\mathbf{S}$-matrix of a linear, non-magnetic and time-independent system is symmetric\cite{Jalas2013}. It  implies that the linear \PT-symmetric Bragg grating (PTBG), studied in this subsection, has the same transmittance for left and right incidence. As such the transmittances are denoted only as transmittance $T$ and are shown in the top panel of Fig. \ref{fig:05_ptbg_t_r_evolve}. The reflectances, however, are different for left and right incidence and are denoted by $R_L$ and $R_R$, respectively, and displayed in the middle and bottom row of Fig. \ref{fig:05_ptbg_t_r_evolve}.
	  
	The transmittance and reflectance of a passive grating has a pronounced gap around the Bragg frequency $f_B$ as a result of collective scattering between high and low refractive index layers. As the gain/loss parameter $n''$ is introduced to the system, the reflectance for left incidence differs from the reflectance for right incidence. Looking at the transmittance spectra, in the top panel of Fig. \ref{fig:05_ptbg_t_r_evolve}, it can be seen that as the gain/loss parameter $n''$ increases the transmission band-gap reduces and almost \textit{unitary transmission}   \index{unitary transmission} $(T=1)$ occurs at $n''=0.02$, with a further increase of $n''$ from this point leading to  \textit{super-unitary} transmission \index{super-unitary transmission} ($T>1$) spectra.
	
	The reflectance for the left incident wave $R_L$ is shown in the middle panel of Fig. \ref{fig:05_ptbg_t_r_evolve}. It can be seen that as the gain/loss parameter $n''$ increases, the reflectance from the left side $R_L$ also increases. Meanwhile, the reflectance for the right incident wave $R_R$ behaves differently in that as the gain/loss parameter increases the right reflectance decreases, and it reaches almost no reflection $R_R=0$ for all frequencies at $n''=0.02$. Operation for gain/loss parameters above this point leads to $R_R>0$.  
	
	It is important to note that for the particular gain/loss parameter of $n''=\Delta n' = 0.02$, the transmittance is unity $(T=1)$ for all frequencies, and the grating is almost reflectionless for the right incident wave $(R_R=0)$ whilst the left incident wave experiences amplified reflection $(R_L>1)$. This particular operation at $n''=\Delta n'$ \cite{Lin2011,Jones2012} is also known as unidirectional invisible operation, since the PTBG is invisible when it is excited from one side (right) but not the other (left). 
	
	Finally, consider the transmittance $T$ and reflectances ($R_L$ and $R_R$) for the case when the gain/loss parameter $n''=0.02429$, shown in Fig. \ref{fig:05_ptbg_t_r_evolve}(f). For this particular value of gain/loss parameter, the value of $T$, $R_L$ and $R_R$ approach infinity at the Bragg frequency $f=f_B$. This particular singularity at $f=f_B$ is associated with the simultaneous coherent perfect absorber-lasing \index{simultaneous coherent-perfect-absorber-lasing} (CPAL) operation point. Once the PTBG enters a lasing state, operating at or above the CPAL point, the system is in an unstable regime since the power inside the structure is increasing exponentially. 
	
	\begin{figure}[t!]	
		\centering
		\begin{overpic}[width=0.8\textwidth,tics=5]{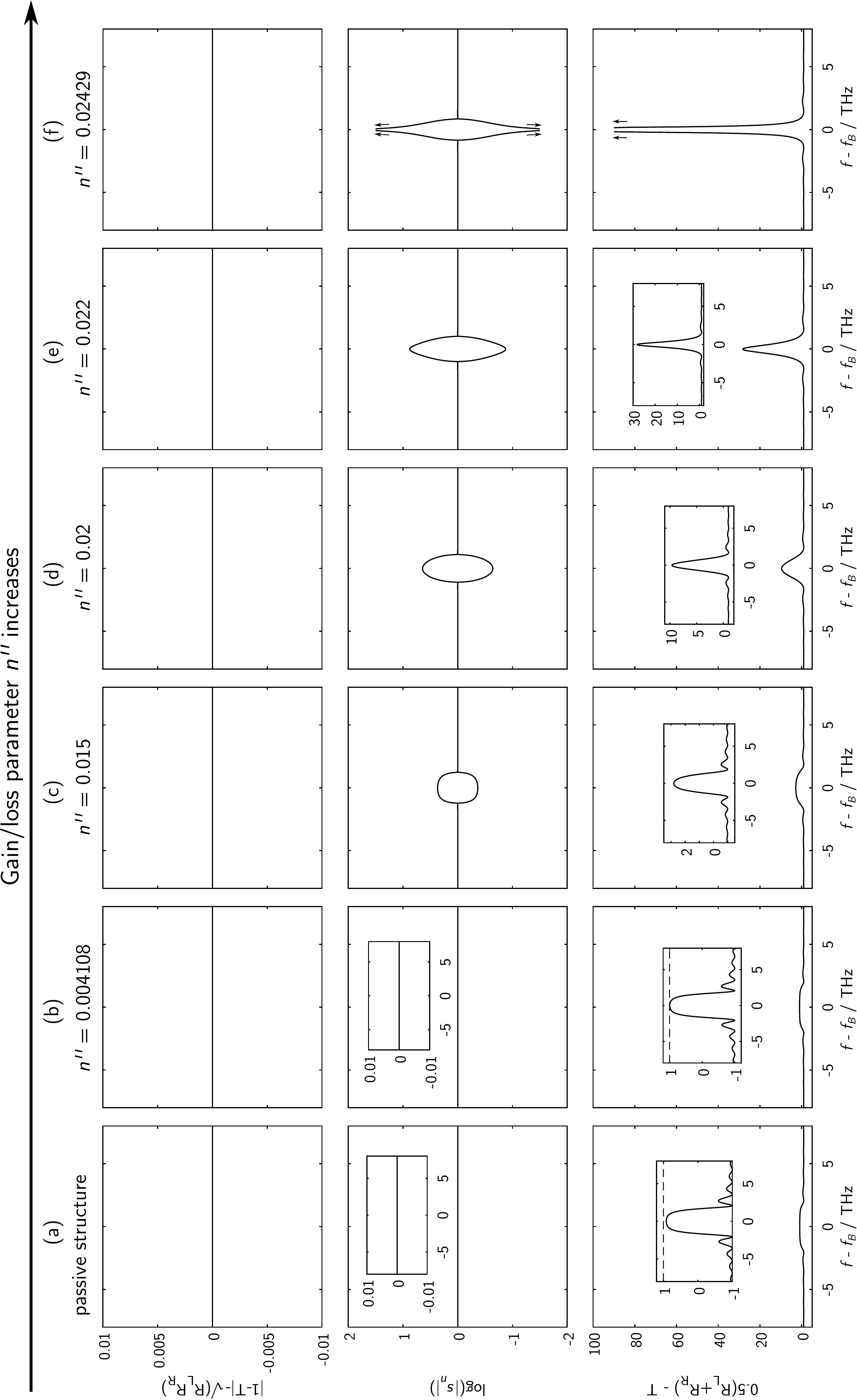}
		\end{overpic}
		\caption[Spectral behaviour of $\mathcal{PT}$-symmetric Bragg grating.]{ {Spectral behaviour of} $ {\mathcal{PT}}$ {-symmetric Bragg grating. (Top panel) the difference between the left and right terms of the general conservation relations. (Middle panel) The magnitude of the eigenvalue of the $\mathbf{S}$-matrix. (Bottom panel) the \PT-phase transition criterion.}}
		\label{fig:05_ptbg_eigenval}
	\end{figure}
	
	In Subsection \ref{eq:genconsrel}, it has been shown that in a \PT-symmetric scattering system a more general conservation relationship, see Eq. (\ref{eq:genconsvrel}), which relates both asymmetric left and right responses is applied. This is reproduced again here: 
	\begin{align}
	|1-T|=\sqrt{R_LR_R} 
	\label{eq:generalconserv}
	\end{align}
	In order to show the validity of Eq. (\ref{eq:generalconserv}), the top row of Fig. \ref{fig:05_ptbg_eigenval} depicts the difference between the left and the right hand side of Eq. (\ref{eq:generalconserv}). It can be seen from the top panel of Fig. \ref{fig:05_ptbg_eigenval} that the difference is zero throughout the spectra for an increasing value of gain/loss parameter $n''$ which implies that the general conservation relation is satisfied in a \PT-symmetric Bragg grating structure. 
	
	Moreover, it is also discussed in Subsection \ref{sec:pttransition} that a \PT-symmetric scattering system may undergo a spontaneous symmetry breaking which could be observed by the magnitude of the eigenvalues of the $\mathbf{S}$-matrix or by a simpler condition defined by the \PT-symmetry transition criterion, given in Eq. (\ref{eq:ptcriterion}). Consider the middle panel of Fig. \ref{fig:05_ptbg_eigenval}. This part of the figure shows the  magnitude of the eigenvalue of the $\mathbf{S}$-matrix, denoted by $|s_{n}|$ where $n\in\{1,2\}$, and bottom panel depicts the \PT-symmetry transition criterion of Eq. (\ref{eq:ptcriterion}) as a function of frequency for different gain/loss parameter $n''$. 
	
	Now consider the middle panel of Fig. \ref{fig:05_ptbg_eigenval} which shows the magnitude of the eigenvalues of the $\mathbf{S}$-matrix, i.e. $|s_n|$ on a semi-log scale for different gain/loss parameters $n''$. It can be seen from the middle panel of Fig. \ref{fig:05_ptbg_eigenval}(a) that the eigenvalues of the passive grating are unimodular $|s_{1,2}|=1$ throughout the frequency spectrum, implying that the $\mathbf{S}$-matrix is orthogonal. However, as the gain/loss is introduced into the system the $\mathbf{S}$-matrix is no longer Hermitian but will be in either the \PT-symmetry or \PT-broken-symmetry phase. As such, in the \PT-symmetry phase the eigenvalue is unimodular $|s_{1,2}|=1$ whilst in the \PT-broken-symmetry phase the \textit{product of the eigenvalues} is unimodular, i.e. $s^*_{1,2}s_{2,1}=1$. Therefore it can be seen from Fig. \ref{fig:05_ptbg_eigenval}(b) that for a gain/loss parameter value of $n''=0.004108$, the PTBG operates in the \PT-symmetric phase throughout the frequency range considered. However from Fig. \ref{fig:05_ptbg_eigenval}(c-f) it can be observed that for larger values of $n''$, the PTBG could operate under the \PT-symmetry phase \index{\PT-symmetric phase} and \PT-broken-symmetry phase \index{\PT-broken symmetry phase}, depending on the operational frequency $f$. It is important to note that since the coupling between the forward and backward propagating waves is strongest at the Bragg frequency $f_B$, the \PT-symmetry will be firstly broken at the Bragg frequency \index{Bragg frequency} and then spread within the band-gap \index{band-gap} of the grating.      
	
	The transition from the \PT-symmetric phase to the \PT-broken-symmetry can be clearly observed by studying the \PT-transition criterion of Eq. (\ref{eq:ptcriterion}). For the operation in the \PT-symmetric phase the criterion is $\frac{1}{2}(R_L+R_R)-T<1$ whilst for operation in the broken-symmetry phase \index{broken-symmetry phase} the criterion is $\frac{1}{2}(R_L+R_R)-T>1$. It is noted that operation at the criterion of $\frac{1}{2}(R_L+R_R)-T=1$ is denoted by the \PT-symmetric breaking point operation. 
	
	The bottom panel of Fig. \ref{fig:05_ptbg_eigenval} plots the criterion $\frac{1}{2}(R_L+R_R)-T$ for different values of gain/loss parameter $n''$; the insets depict the detail of the criterion value with the dashed line denoting the $\frac{1}{2}(R_L+R_R)-T=1$ value. It can be seen that for the passive grating structure, the criterion value is below 1 throughout the frequency spectrum. As the gain/loss parameter value increases, the criterion value also increases. It is noted that at the particular value of $n''=0.004108$, the criterion value at the Bragg frequency \index{Bragg frequency} $f_B$ is just touching the dashed line. It implies that the value of gain/loss parameter $n''=0.004108$ indicates the initial \PT-symmetry breaking which occurs at the Bragg frequency and which is followed by other frequencies within the band-gap region proximity.
	
	It is important to inspect the operation at the gain/loss parameter value of $n''=0.02429$, depicted by Fig. \ref{fig:05_ptbg_eigenval}(f). Although the PTBG exhibits CPAL operation, the \PT-system satisfies the general conservation relationship. Furthermore, the eigenvalues of the $\mathbf{S}$-matrix show a strong singularity at the Bragg frequency $f_B$ with one of the eigenvalues approaching infinity whilst the other approaches zero. This implies that the structure supports both lasing and coherent-perfect-absorber operation simultaneously at the same operating frequency and at the same gain/loss parameter value. The singularity signatures are also observed in Fig. \ref{fig:05_ptbg_eigenval}(f, bottom panel) where the criterion at the Bragg frequency is also approaching infinity.                
	
	\begin{figure}[t]
		\centering
		\begin{overpic}[width=0.5\textwidth,tics=5] {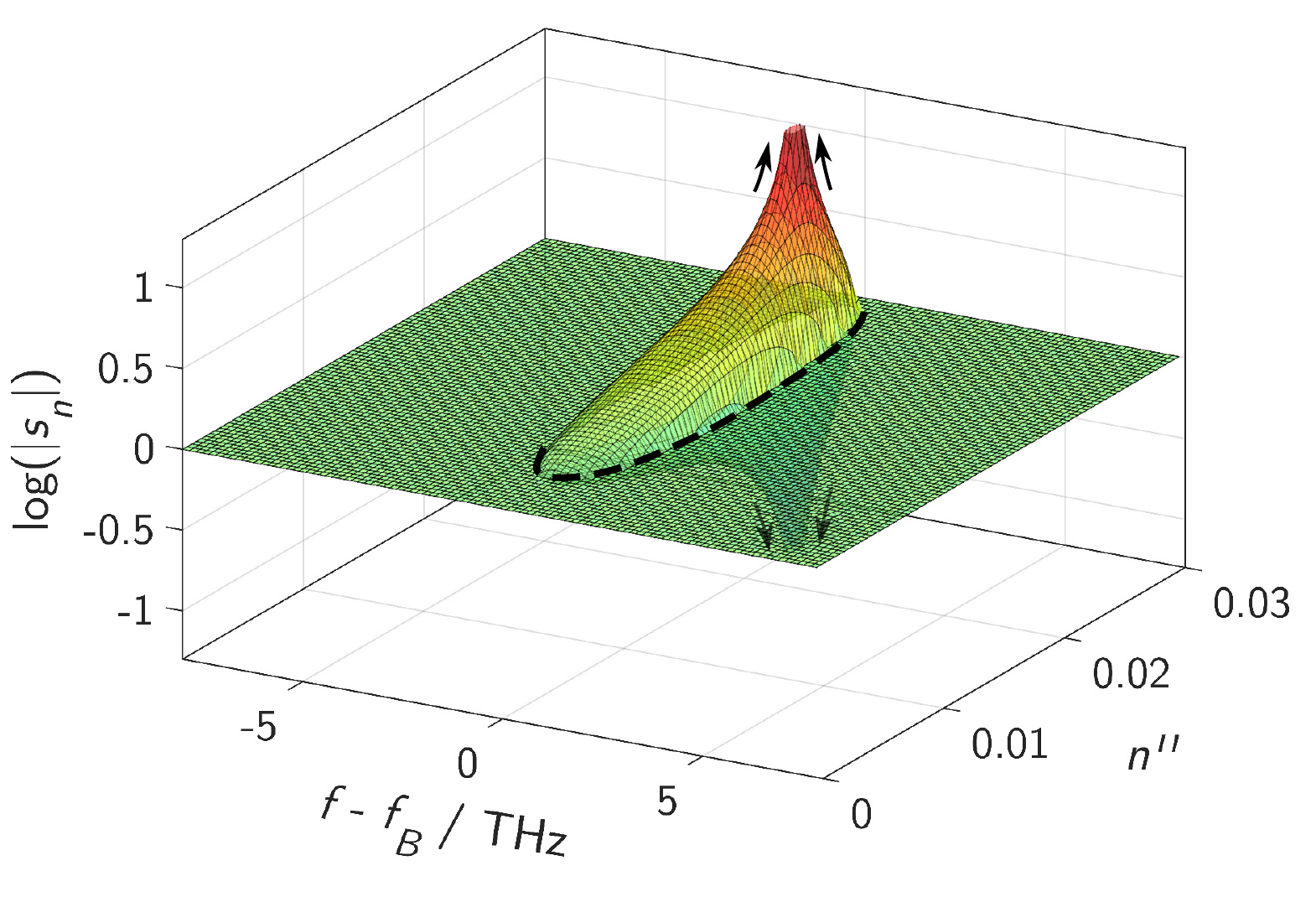}
		\end{overpic}
		\caption[Magnitude of the eigenvalues of the $\mathbf{S}$-matrix]{ {Magnitude of the eigenvalues of the} $\mathbf{S}${-matrix. Plotted on a semi-log scale as a function of frequency and for different gain/loss parameter $n''$}}
		\label{fig:05_eigenvalue_3d}
	\end{figure}
	
	In order to study the \PT-symmetry phase transition \index{phase transition} process, Fig. \ref{fig:05_eigenvalue_3d} depicts the magnitude of the eigenvalues of the $\mathbf{S}$-matrix on a semi-log scale as a function of both gain/loss parameter and the operating frequency $f$. For this figure, the PTBG considered is the same as that presented in Fig. \ref{fig:05_ptbg_t_r_evolve} and Fig. \ref{fig:05_ptbg_eigenval}. As a passive grating, i.e. with no gain/loss, the eigenvalues are uni-modular (see Fig. \ref{fig:05_eigenvalue_3d} for $n''=0$). However, as the gain/loss increases, the \PT-symmetry starts to break which initially happens at $f_B$ for the gain/loss parameter value of $n''=0.004108$. For a further increase of gain/loss in the system, more \PT-symmetry breaking is observed. The black dashed line in this figure denotes the value of $\frac{1}{2}(R_L+R_R)-T=1$, i.e. the \PT-symmetry breaking point \index{\PT-symmetry breaking point}. Furthermore this figure demonstrates that \PT-symmetry breaking occurs at a frequency located within the band-gap region of the grating. It can be explained since propagation of a wave at a frequency outside of the band-gap experiences almost no dispersion, and the interaction between the forward and the backward propagation wave is negligible. Moreover, this figure also shows the singularity point of the eigenvalues of the $\mathbf{S}$-matrix, with one eigenvalue approaching infinity whilst the other approaches zero. Operation at this singular point is associated with the CPAL point which is depicted in more detail in Fig. \ref{fig:05_ptbg_eigenval}(f). It is important to note that in practice as the structure reaches the CPAL point the system becomes unstable as it is now operating as a laser cavity, hence operation beyond the CPAL point leads to unstable operation.
	
\section{Modelling Parity-Time (\PT) Symmetric Bragg Grating with a Realistic Gain/loss Material Model}
	The \PT-symmetric Bragg grating \index{\PT-symmetric Bragg grating} (PTBG) studied so far has been considering a simple non-dispersive gain/loss material model. This section demonstrates the impact of a realistic gain properties, dispersive and saturable, on the performance of a PTBG. For that reason, this section will first describe the dispersive and saturable gain model used and the implementation of such a model within the time-domain Transmission-Line Modelling \index{Transmission-Line Modelling} (TLM) method.

	\subsection{Time-Domain Modelling of Dispersive and Saturable Gain} \label{sec:realisticgain}
	Light amplification phenomena can be explained using the concept of energy levels \index{energy levels} and the transitions of electrons between the energy levels\cite{Yariv1989,Iizuka2002}. To elaborate the concept of gain by the mechanism of electron transition between different energy levels consider a three-energy-level model. The three-energy-level model is a simplified model  which is typically used to described light amplification phenomena in an erbium-doped based optical amplifier \cite{Saleh2007,Liu2005,Iizuka2002}. In such a configuration, there are three energy levels denoted by $E_1$, $E_2$ and $E_3$, with $E_1$ being the lowest energy level. Light amplification occurs by a stimulated emission process of electron transition from $E_2$ to $E_1$.   
	
	For the case when the electron transition between $E_2$ and $E_1$ is considered to be \textit{homogeneous}, the electron response to the incoming light is characterised by the same atomic transitional angular frequency $\omega_\sigma$ and the same time relaxation \index{time relaxation} parameter, $\tau$. In such an homogeneous system, the time relaxation parameter $\tau$ models the time required by the electrons to rest after transition. The finite material response introduces a broadening in the spectrum of the emitted light in the shape of Lorentzian lineshape \index{Lorentzian lineshape} function. A macroscopic model of the homogeneously broadened gain medium can conveniently be modelled through the electrical conductance of the medium as\cite{Hagness1996},                 
	\begin{align}
	\sigma_e(I,\omega) = \mathbb{S}(I) \frac{\sigma_0}{2}   \left[
	\frac{1}{1+j(\omega-\omega_\sigma)\tau} +
	\frac{1}{1+j(\omega+\omega_\sigma)\tau}
	\right]  
	\label{eq:causalgain}
	\end{align}
	where, $\omega_\sigma$ denotes the atomic transitional angular frequency, $\tau$ is the atomic relaxation time parameter, and $\sigma_0$ is related to the conductivity peak value that is set by the pumping level at $\omega_\sigma$. The saturation coefficient \gainsat is non-linear in nature as a consequence of the finite number of electrons available in the case of large incident signal and is conveniently described as\cite{Liu2005,Hagness1996,Siegman1986}, 
	\begin{align}
	\mathbb{S}(I) =  \frac{1}{1+(I/I_{sat})} 
	\end{align}
	In the case of a small incident signal the saturation coefficient is typically negligible. Note that value of the saturation intensity \index{saturation intensity} $I_{sat}$ is dependent on the details of structure and treatment of the material\cite{mich2015}.
	
	Overall, the frequency domain relative dielectric permittivity is given by, 
	\begin{align}
	\varepsilon_r(\omega,I)=  1+ \chi_e(\omega) -
	j\mathbb{S}(I)\frac{\sigma_0}{2\varepsilon_0\omega} 
	\left[ \frac{1}{1+j(\omega-\omega_\sigma)\tau} + \frac{1}{1+j(\omega+\omega_\sigma)\tau}
	\right]  
	\label{eq:gainmod1}
	\end{align} 
	It is important to note that the material model given by Eq. (\ref{eq:gainmod1}) satisfies the Kramers-Kronig relations \index{Kramers-Kronig relations} by the fact that a change in the imaginary part causes the real part of the dielectric constant to be dispersive and it meets the analytic condition of the Fourier transform, i.e. all singularities of the model are located in the upper half-plane of the complex frequency plane\cite{Landau1984,Zyablovsky2014}. 
	
	In order to associate the conductivity model given in Eq. (\ref{eq:gainmod1}) with the resulting gain, assume that the dielectric susceptibility is constant and real, i.e. $\chi_e(\omega)=\chi_e$ and consider small signal gain. The relative permittivity can be simplified as, 
	\begin{align}
	\varepsilon_r(\omega) = 1+\chi_e+ \frac{\sigma_e''(\omega)}{\varepsilon_0 \omega} -j\frac{\sigma_e'(\omega)}{\varepsilon_0 \omega} 
	\label{eq:simpdielcons}
	\end{align}  
	where the frequency domain (small signal) conductivity has been considered in the form of $\sigma_e(\omega)=\sigma_e'(\omega)+j\sigma_e''(\omega)$, so that  the real and imaginary parts of the conductivity are given by,
	\begin{equation}
	\begin{split}
	\sigma_e'(\omega) &= \sigma_0 \frac{  1 + (\omega_\sigma^2+\omega^2)\tau^2}{\{1 + (\omega_\sigma^2-\omega^2)\tau^2\}^2 + 4\omega^2\tau^2}  \\
	\sigma_e''(\omega) &= \sigma_0 \frac{  (\omega\tau) \{-1 + (\omega_\sigma^2-\omega^2)\tau^2\}   }{\{1 + (\omega_\sigma^2-\omega^2)\tau^2\}^2 + 4\omega^2\tau^2}  
	\end{split}
	\label{eq:realimagcond}
	\end{equation}
	
	In the refractive index formalism, the propagation constant can also be expressed as, 
	\begin{align}
	\gamma=\alpha+j\beta=j\frac{\omega}{c_0} n(\omega)
	\label{eq:propcst}
	\end{align}
	where the complex frequency-domain refractive index is defined as,
	\begin{align}
	n(\omega)\equiv& n'(\omega)+jn''(\omega)=\sqrt{\varepsilon_r(\omega)}
	\label{eq:refracindexgain}
	\end{align} 
	Consequently, the phase constant ($\beta$) and gain ($\alpha$) depend only on the real and imaginary parts of the refractive index respectively as, 
	\begin{align}
	\alpha &= -\frac{\omega}{c_0} n''(\omega)  \label{eq:gain}\\
	\beta &= \frac{\omega}{c_0}n'(\omega) 
	\end{align}   
	By assuming propagation in the $+z$ direction as $\e^{-\gamma z}$, it can be seen from Eq. (\ref{eq:realimagcond}), Eq. (\ref{eq:propcst}) and Eq. (\ref{eq:gain}) that gain is achieved by having $\sigma_0<0$. 
	
	It is also important to note that the three-level system also describes light absorption phenomena since, in the absence of external pumping, most of the electrons are at $E_1$ and the incoming light signal induces an upward transition from $E_1$ to $E_2$. The upward transition induces loss at the frequency  corresponding  to the appropriate energy, $E_2-E_1$. Mathematically, this induced absorption loss can be modelled by Eq. (\ref{eq:gainmod1}) by having $\sigma_0>0$.
	
	The realistic gain/loss material model Eq. (\ref{eq:gainmod1}) is now implemented within the time-domain Transmission-Line modelling \index{Transmission-Line Modelling} (TLM) method in one-dimension. The TLM method is a flexible time-stepping numerical technique that has been extensively characterised and used over many years \cite{Hoefer1985,Christopoulos1995}. The TLM method is based upon the analogy between the propagating electromagnetic fields and voltages impulses travelling in an interconnected mesh of transmission-lines. Successive repetitions of a scatter-propagate procedure \index{scatter-propagate procedure} provide an explicit and stable time-stepping algorithm that mimics electromagnetic field behaviour to second-order accuracy in both time and space \cite{Hoefer1985,Christopoulos1995}. In this chapter, we do not attempt to describe the basic of TLM method itself and readers are referred to some excellent references\cite{Christopoulos1995,Paul1998}. However, in this chapter, an alternative TLM formulation using a bilinear $\mathcal{Z}$-transformation \index{$\mathcal{Z}$-transform} of Maxwell's equations approach is employed \cite{Paul2002,Paul1998,Paul1999a}. In this approach, the TLM is formulated using less of an electical analogy and more of transmission-line characteristics and a $\mathcal{Z}$-transformation of Maxwell's equations\cite{Paul2002,Paul1998,Paul1999a}. In particular, this approach offers flexibility in the implementation of dispersive and non-linear material properties\cite{Janyani2004a,Janyani2005a,Paul2002,Paul1998,Paul1999a}.   
	
	Without losing generality in this chapter, the implementation for a one-dimensional problem is considered, implementation for two-and three-dimensions follows similarly and the two-dimensional implementation is presented in the following chapter. For that reason, we shall consider Maxwell's equations for a one-dimensional problem with the electric field polarisation in the $y$-direction as,
	\begin{align}
	-\frac{\partial}{\partial x}
	\begin{bmatrix}
	H_z \\
	E_y
	\end{bmatrix} =
	\begin{bmatrix}
	\sigma_e \ast E_y \\
	0
	\end{bmatrix} + \frac{\partial}{\partial t} 
	\begin{bmatrix}
	\varepsilon_0(E_y + \chi_e \ast E_y) \\
	\mu_0 H_z
	\end{bmatrix} 
	\label{eq:max1d}
	\end{align}
	In Eq. (\ref{eq:max1d}), the curl Maxwell's equations are displayed in a compact matrix notation, where $\ast$ denotes the time-domain convolution operator. Maxwell's equation Eq. (\ref{eq:max1d}) can be expressed in circuit format by utilising the field-circuit equivalences \cite{Christopoulos1995,Paul1999a}, which are summarised in  Table \ref{tab:emtlequi}. The circuit form of Eq. (\ref{eq:max1d}) is given as,
	\begin{align}
	-\frac{\partial}{\partial x} \Delta x
	\begin{bmatrix}
	I_z \\
	V_y
	\end{bmatrix} 
	=
	\begin{bmatrix}
	G_e \ast V_y \\
	0
	\end{bmatrix} + \frac{\partial}{\partial t} 
	\begin{bmatrix}
	C_0\left( V_y + \chi_e \ast V_y \right)  \\
	L_0I_z
	\end{bmatrix} 
	\label{eq:1dtlmform}
	\end{align}
	By introducing the following normalisation transformation,
	\begin{align}
	\begin{split}
	x \rightarrow X \Delta x  &\quad | \quad \partial x \rightarrow \Delta x \partial X \\ 
	t \rightarrow T \Delta t  &\quad | \quad \partial t \rightarrow \Delta t \partial T 
	\end{split}
	\label{eq:spacetimenorm}
	\end{align}
	where $X$ and $T$ are dimensionless variables and $\Delta x$ and $\Delta t$ are discretisation length and the time-stepping parameter, 
	\begin{table*}[t]\centering
		\caption[Equivalences of the field and transmission-line quantities.]{ {Equivalences of the field and transmission-line quantities}\cite{Christopoulos1995,Paul1998}.}
		\begin{tabular}{@{}rccrccc@{}} 
			
			\toprule
			\multicolumn{3}{c}{Field theory} & \multicolumn{3}{c}{Transmission line theory} & \\  
			\cmidrule(r){1-3} \cmidrule(r){4-6}
			Quantity       & Symbol         & unit  & Quantity      & Symbol & unit  & Transformations\\ 
			\midrule
			Electric field & $E$            & [V/m] & Voltage       & $V$    & [V]   & $E\leftrightarrow-\frac{V}{\Delta x}$\\
			Magnetic field & $H$            & [A/m] & Current       & $I$    & [A]   & $H\leftrightarrow-\frac{I}{\Delta x}$\\
			Permittivity   & $\varepsilon$  & [F/m] & Capacitance   & $C$    & [F]   & $\varepsilon\leftrightarrow\frac{C}{\Delta x}$\\
			Permeability   & $\mu_0$        & [H/m] & Inductance    & $L$    & [H]   & $\mu_0\leftrightarrow\frac{L}{\Delta x}$\\
			Conductivity   & $\sigma$       & [S/m] & Conductivity  & $G_e$  & [S]   & $\sigma\leftrightarrow\frac{G_e}{\Delta x}$\\
			\bottomrule
			
		\end{tabular}
		\label{tab:emtlequi}
	\end{table*} 
	equation Eq. (\ref{eq:1dtlmform}) can be simplified further as a single unit operation (volt) as, 
	\begin{align}
	-\frac{\partial}{\partial X}
	\begin{bmatrix}
	i_z \\
	V_y
	\end{bmatrix} =
	\begin{bmatrix}
	g_e * V_y \\
	0
	\end{bmatrix} + 
	\frac{\partial}{\partial T}
	\begin{bmatrix}
	V_y  \\
	i_z
	\end{bmatrix} +
	\frac{\partial}{\partial T}
	\begin{bmatrix}
	\chi_e * V_y \\
	0
	\end{bmatrix} 
	\label{eq:1dtlm1}
	\end{align}
	Here, the normalised conductivity and current parameters are defined as $g_e=G_e Z_\text{TL}$ and $i_z = I_z Z_\text{TL}$, where $Z_\text{TL}$ denotes the characteristic impedance of the transmission-line and has been adopted to correspond to the properties in free-space, hence $Z_\text{TL}=\sqrt{L_0/C_0}$ and $\Delta x = c_0\Delta t$ where $c_0=1/\sqrt{\varepsilon_0\mu_0}$.
	
	By utilising the travelling-wave format \index{travelling-wave format} \cite{Paul2002,Paul1998,Paul1999a}, 
	\begin{align}
	-\dfrac{\partial i_z}{\partial X} -\dfrac{\partial V_y}{\partial T} &= 2V_4^i+2V_5^i -2V_y  \\
	-\dfrac{\partial V_y}{\partial X} -\dfrac{\partial i_z}{\partial T} &= 2V_4^i -2V_5^i -2i_y 
	\end{align}
	where $V_4^i$ and $V_5^i$ denote the incident impulses coming from the left and right respectively. The travelling-wave form of Eq. (\ref{eq:1dtlm1}) in the Laplace-domain is given as, 
	\begin{align}
	2
	\begin{bmatrix}
	V_y^r \\
	i_z^r
	\end{bmatrix} \equiv
	2
	\begin{bmatrix}
	V_4^i+V_5^i \\
	V_4^i-V_5^i
	\end{bmatrix} =
	2\begin{bmatrix}
	V_y \\
	i_z
	\end{bmatrix}+
	\begin{bmatrix}
	g_e V_y \\
	0
	\end{bmatrix} + \bar{s} 
	\begin{bmatrix}
	p_{ey} \\
	0
	\end{bmatrix} 
	\label{eq:1dtlm2}
	\end{align}
	In Eq. (\ref{eq:1dtlm2}), the convolution operator $*$, which appeared in Eq. (\ref{eq:1dtlm1}), has been transformed to a simple multiplication in the frequency domain and $p_{ey}=\chi_eV_y$ denotes the normalised dielectric polarisation. Note that the normalised Laplacian operator \index{Laplacian operator} is $\bar{s} = \partial/\partial T$. Performing a bilinear $\mathcal{Z}$-transform \index{$\mathcal{Z}$-transform} of the normalised Laplacian operation \cite{Paul2002,Paul1998,Paul1999a},
	\begin{align}
	\bar{s} \xrightarrow{\mathcal{Z}} 2\left( \frac{1-z^{-1}}{1+z^{-1}}\right)  
	\label{eq:normz}
	\end{align}
	equation (\ref{eq:1dtlm2}) becomes, in the $\mathcal{Z}$-domain, 
	\begin{align}
	2
	\begin{bmatrix}
	V_y^r \\
	i_z^r
	\end{bmatrix} =
	2\begin{bmatrix}
	V_y \\
	i_z
	\end{bmatrix}+
	\begin{bmatrix}
	g_e  V_y \\
	0
	\end{bmatrix} + 2\left( \frac{1-z^{-1}}{1+z^{-1}}\right) 
	\begin{bmatrix}
	p_{ey} \\
	0
	\end{bmatrix} 
	\label{eq:1dtlmfinal}
	\end{align} 
	Equation (\ref{eq:1dtlmfinal}) is suited for material modelling with dispersive and non-linear properties, which are modelled through the dielectric polarisation $p_{ey}$ and the conductivity $g_e$. Right after obtaining the voltage $V_y$ and current $i_z$ quantities, the new scattered voltage impulses can obtained by\cite{Christopoulos1995,Paul2002,Paul1998,Paul1999a}, 
	\begin{align}
	\begin{split}
	V_4^r &= V_y - V_4^i \\
	V_5^r &= V_y - V_5^i 
	\end{split} 
	\label{eq:1dnewscatt}
	\end{align}
	and communicated to the neighbouring nodes during the connection process.
	
	Now, a digital filter \index{digital filter} based on the material model Eq. (\ref{eq:gainmod1}) is developed. The purpose of designing a digital filter of the material model is to facilitate the implementation of the frequency-domain material model within the time-domain TLM method Eq. (\ref{eq:1dtlm2}). The gain (or loss) material model given in Eq. (\ref{eq:causalgain}) can be conveniently expressed in the Laplace domain as,  
	\begin{align}
	\sigma_e(I,s) =\mathbb{S}(I) \sigma_0 
	\left[
	\frac{K_1s+(K_1)^2}{s^2 +2K_1s+(K_2)^2}
	\right]   
	\label{eq:gainlap1}
	\end{align}
	where the constants $K_1$ and $K_2$ are defined as,
	\begin{align*}
	K_1 = \frac{1}{\tau} \quad \mbox{and} \quad K_2 = \frac{1+(\omega_\sigma\tau)^2}{\tau^2} 
	\end{align*}
	Using the normalisation procedure introduced previously, the material model of Eq. (\ref{eq:gainlap1}) in the TLM form can be expressed as,  
	\begin{align}
	g_e(I,s) =\mathbb{S}(I) g_0 
	\left[
	\frac{K_1s+(K_1)^2}{s^2 +2K_1s+(K_2)^2}
	\right]  
	\label{eq:ge1} 
	\end{align}
	and by performing the bilinear $\mathcal{Z}$-transformation \index{$\mathcal{Z}$-transform} on the Laplacian operator \index{Laplacian operator} as,
	\begin{align}
	g_e(I,z) =\mathbb{S}(I) g_0 
	\left[
	\frac{K_3+z^{-1}(K_4)+z^{-2}(K_5)}{K_6 ++z^{-1}(K_7)++z^{-2}(K_8)}
	\right]  
	\label{eq:ge2}  
	\end{align}
	where, the constants in Eq. (\ref{eq:ge2}) are given by, 
	\begin{align}
	K_3 = 2K_1\Delta t + (K_1\Delta t)^2 &\quad \mbox{;} \quad K_4 =  2(K_1\Delta t)^2 \\
	K_5 = -2K_1\Delta t + (K_1\Delta t)^2 &\quad \mbox{;} \quad K_6 = 4 + 4 K_1\Delta t + K_2 (\Delta t)^2 \\
	K_7 = -8 + 2K_2(\Delta t)^2 &\quad \mbox{;} \quad K_8 =  4 - 4 K_1\Delta t + K_2 (\Delta t)^2
	\end{align}
	Furthermore, any system with a causal response \index{causal response} can always be described as a feedback system whose current response depends on a past event. Hence it can be shown that \cite{Paul1999a} ,   
	\begin{align}
	(1+z^{-1})g_e = g_{e0} + z^{-1}(g_{e1} + \bar{g}_e(z)) 
	\label{eq:ge3} 
	\end{align}
	where the constants $g_{e0}$ and $g_{e1}$ and the causal response $\bar{g}_e(z)$ are given by, 
	\begin{align}
	\begin{split}
	g_{e0} = g_s \left( \frac{K_3}{K_6}\right), \quad
	g_{e1} = 0 , \quad
	\bar{g}_e(z) = \frac{b_0 + z^{-1}b_1 + z^{-2}b_2}{1 - z^{-1}(-a_1) - z^{-2}(-a_2)} 
	\end{split}
	\label{eq:gaincomponents}
	\end{align}
	with the corresponding constants defined as,
	\begin{align}
	\begin{split}
	& g_s = g_0 \left( \dfrac{1}{1+(I/I_{sat})}\right);  \quad b_0 = g_s \left( \dfrac{K_3}{K_6}\right)   
	\left( \dfrac{K_3+K_4}{K_3}  - \dfrac{K_7}{K_6} \right)  \\
	& b_1 = g_s \left( \dfrac{K_3}{K_6}\right)   \left( \dfrac{K_4+K_5}{K_3}  - \dfrac{K_8}{K_6} \right); \quad 
	b_2 = g_s \left( \dfrac{K_3}{K_6}\right)   \left( \dfrac{K_5}{K_3}  \right)  \\
	& a_1 =  \dfrac{K_7}{K_6}; \quad
	 a_2 =  \dfrac{K_8}{K_6}.
	\end{split}
	\end{align} 
	
	We are now ready to implement the digital filter \index{digital filter} for gain (or loss) material Eq. (\ref{eq:ge3}) within the 1D-TLM method. For convenience, the first row of Eq. (\ref{eq:1dtlmfinal}) is reproduced here, 
	\begin{align}
	2V_y^r = 2V_y + g_eV_y +2 \left( \frac{1-z^{-1}}{1+z^{-1}} \right) p_{ey}
	\label{eq:1dgainmod1}
	\end{align}  
	After multiplying both sides by $(1+z^{-1})$ and rearranging, Eq. (\ref{eq:1dgainmod1}) can also be expressed as,  
	\begin{align}
	(2V_y^r-2V_y)+z^{-1}(2V_y^r-2V_y)  = (1+z^{-1})g_eV_y + 2(1-z^{-1})p_{ey}
	\label{eq:1dgainmod2}
	\end{align}
	Substituting the digital filter for the conductivity given in Eq. (\ref{eq:ge3}), and by further assuming the case of linear and dispersionless dielectric polarisation $p_{ey} = \chi_{e\infty} V_y$, Eq. (\ref{eq:1dgainmod2}) reduces to  
	\begin{align}
	2V_y^r + z^{-1}(S_{ey}) = K_{e2}V_y 
	\label{eq:1dgaintlm}
	\end{align} 
	where the accumulative past response is given by,
	\begin{subequations}
		\begin{align}
		S_{ey} &{}= 2V_y^r + K_{e1}V_y +S_{ec}  \\
		S_{ec} &{}= -\bar{g}_eV_y 	
		\end{align}
	\end{subequations}
	with the constants $K_{e1}$ and $K_{e2}$ defined as,  
	\begin{subequations}
		\begin{align}
		K_{e1} &{}= -(2+g_{e1}-2\chi_{e\infty})  \\
		K_{e2} &{}= 2+g_{e0}-2\chi_{e\infty} 	
		\end{align}
	\end{subequations}
	and $g_{e0}$, $g_{e1}$ and $\bar{g}_e$ are as in Eq. (\ref{eq:gaincomponents}). 
	
	\begin{figure}[t]
		\begin{overpic}[width=0.7\textwidth,tics=5] {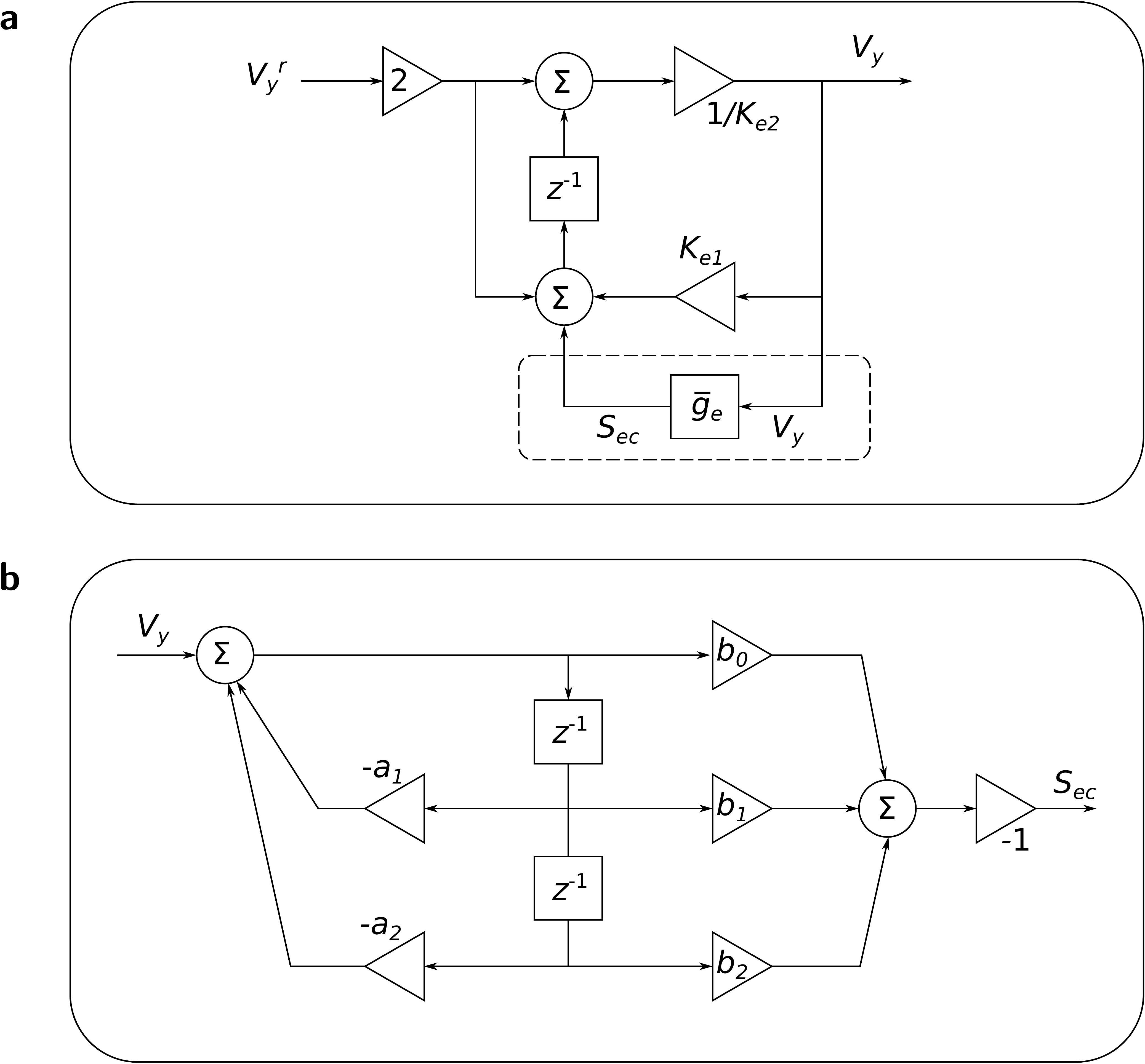}
		\end{overpic}
		\centering
		\caption[Signal flow diagram modelling gain material in TLM algorithm]{ {Signal flow diagram modelling gain material in TLM algorithm. ({a}) Overall signal flow diagram from the incoming voltage impulses $V_y^r$ to the resulting nodes voltage $V_y$. ({b}) Detail field updating scheme of conductivity model of gain material which is marked in the dashed box in ({a}).}}
		\label{fig:03_sigdiag}
	\end{figure}
	
	The signal flow diagram \index{signal flow diagram} of system Eq. (\ref{eq:1dgaintlm}) is illustrated in Fig. \ref{fig:03_sigdiag}(a), the subsystem defining the conductivity digital filter \index{digital filter} system (within the dashed line box) is detailed in Fig. \ref{fig:03_sigdiag}(b). It is also noted here that, for the case of a saturable gain (or loss) model, the saturation coefficient $\mathbb{S}(I)$ is updated as follows: if $|V_y|$ at the time-step $T$ is greater than $|V_y|$ at $T-1$ at the same location in space, then $\mathbb{S}(I)$ is updated using the last value of $|V_y|$. However, if $|V_y|$ has decreased from its previous value, it is not updated; hence $\mathbb{S}(I)$ remains based on the most recent peak value. In this manner, intensity feedback in the time-domain retains, as much as possible, its frequency domain meaning\cite{Hagness1996}. Thus the time-averaged intensity $I$ can be calculated as,
	\begin{align}
	I = \frac{1}{2} \frac{E_y^2}{\eta} = \frac{1}{2} \frac{V_y^2}{\eta\Delta x^2}, \quad \mbox{where} \quad 
	\eta = \frac{\eta_0}{n'}
	\label{eq:intensity1d}
	\end{align}
	where $\eta_0=\sqrt{\mu_0/\varepsilon_0}$ is the free-space impedance of a normally incident wave and $n'$ is the real-part of the refractive index.
	
	\subsection{Impact of Dispersion on the Properties of a \PT-Bragg Grating} \label{sec:impdispptbg}
	Now, consider the PTBG similar to that illustrated in Fig. \ref{fig:05_ptbgillus} with the exception that the gain and loss are defined by the realistic dispersive gain/loss model developed in previous subsection. As such the relative permittivity distribution in a single unit cell, $\bar{\varepsilon}(x)$, along the propagation direction $x$ can be expressed as, 
	\begin{align}
	\bar{\varepsilon}(x,\omega) = 
	\begin{cases}
	\bar{\varepsilon}_b + \Delta \bar{\varepsilon}' -j\dfrac{\sigma_e(\omega)}{\varepsilon_0\omega}  
	,\quad x<\dfrac{\Lambda}{4}  \\
	\bar{\varepsilon}_b - \Delta \bar{\varepsilon}' -j\dfrac{\sigma_e(\omega)}{\varepsilon_0\omega}  
	,\quad \dfrac{\Lambda}{4}<x<\dfrac{\Lambda}{2}\\
	\bar{\varepsilon}_b - \Delta \bar{\varepsilon}' +j\dfrac{\sigma_e(\omega)}{\varepsilon_0\omega}  
	,\quad \dfrac{\Lambda}{2}<x<\dfrac{3\Lambda}{4}\\
	\bar{\varepsilon}_b + \Delta \bar{\varepsilon}' +j\dfrac{\sigma_e(\omega)}{\varepsilon_0\omega} 
	,\quad \dfrac{3\Lambda}{4}<x<\Lambda \\
	\end{cases}
	\label{eq:dielprof}
	\end{align}
	where $\Delta \bar{\varepsilon}'$ denotes the constant modulation of the real part of the dielectric permittivity and $\varepsilon_0$ denotes the free-space permittivity. The material conductivity $\sigma_e$ is a function of frequency as was given in Eq. (\ref{eq:causalgain}). From Eq. (\ref{eq:dielprof}), it can be seen that the first two sections of the PTBG have gain while the other two sections are lossy. Moreover, it is a common practice in optics to denote dielectric material properties using the complex refractive index, $n=n'+jn''$ which is related to the complex dielectric permittivity by $n=\sqrt{\bar{\varepsilon}(\omega)}$. In this subsection, we will investigate the impact of dispersion; for that reason a small incident signal is considered, hence the saturation coefficient is negligible and $\mathbb{S}=1$. Operation with a strong signal will be discussed in the next subsection which will also include the non-linear Kerr effect \index{Kerr effect}. 
	
	For definiteness, consider a PTBG  with the following material parameters: the background dielectric constant $\bar{\varepsilon}_b=(3.625)^2$ and modulation of the real-part of dielectric constant $\Delta \bar{\varepsilon}'=(0.02)^2$ as used in \cite{Phang2014d}. The parameter related to the gain/loss material used is similar to that reported in \cite{Hagness1996}, in which the atomic transition angular frequency $\omega_\sigma=2\pi(336.85)$ rad/ps, and time relaxation parameter $\tau=0.1$ ps. The PTBG is designed as follows: the grating has $N=200$ and the Bragg frequency is centred at the atomic transitional frequency $f_B=336.85$ THz. It follows that the physical length of a unit cell is $\Lambda=112.7$ nm.  
	
	\begin{figure}[]	
		\centering
		\begin{overpic}[width=0.8\textwidth,tics=5]{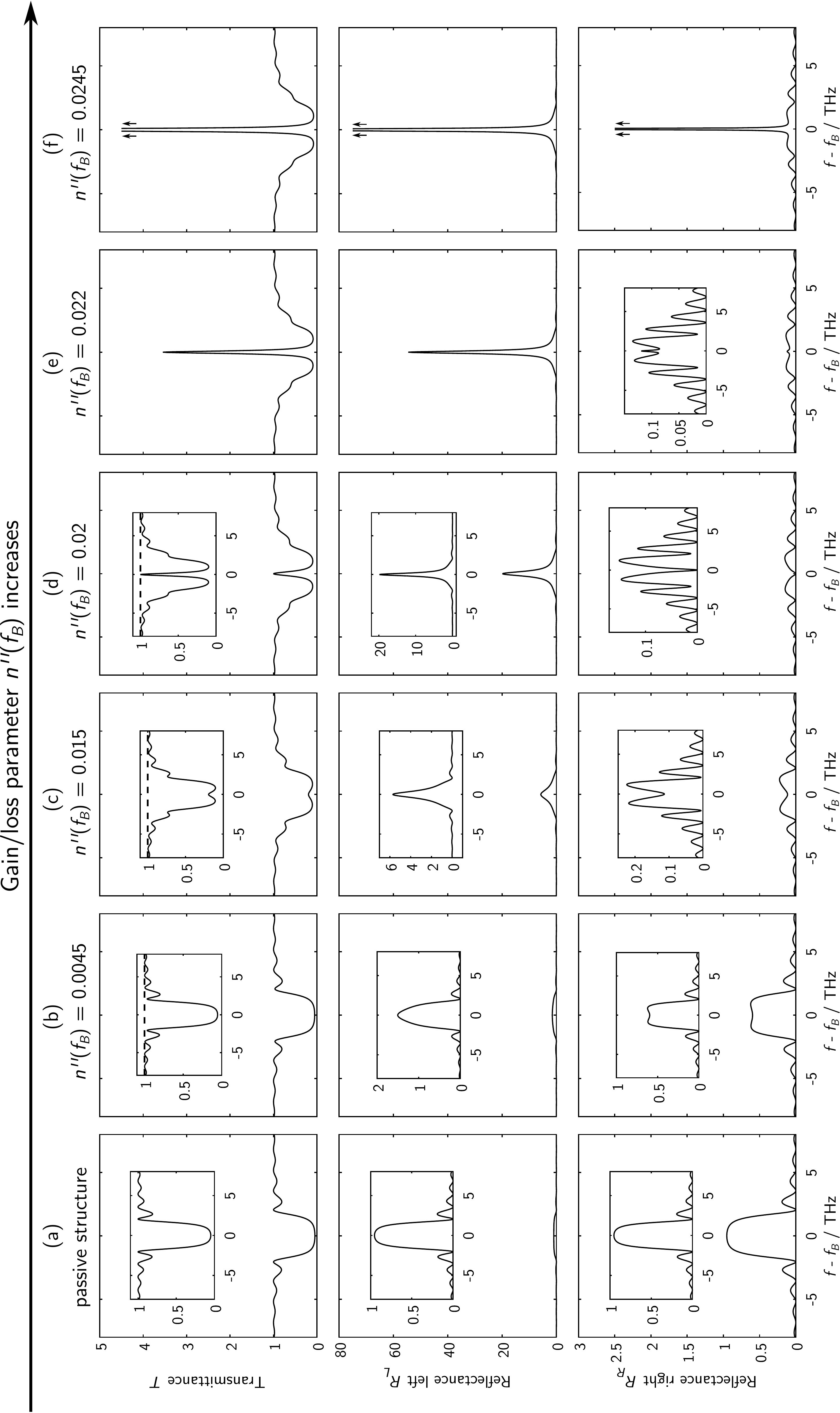}
		\end{overpic}
		\caption[The impact of dispersive gain/loss parameter on the performance of PTBG.]{{The impact of dispersive gain/loss parameter on the performance of PTBG. The transmittance $T$, reflectance for the left $R_L$ and right incident $R_R$ are displayed at the top, middle and bottom panel respectively. Six different value of gain/loss parameter: passive structure,  $n''(f_B)=0.0045$, $0.015$, $0.02$, $0.022$ and $0.0245$ are considered. The inset shows the magnified spectra.}}
		\label{fig:06_ptbg_t_r_evolve_realistic_gl}
	\end{figure}
	
	It is emphasised here that the gain/loss considered in this chapter is different to that in Subsection \ref{sec:noncausalPTBG}. In Subsection \ref{sec:noncausalPTBG} the gain/loss parameter $n''$ is non-dispersive. In this section the gain/loss parameter is dispersive and causal \index{causal response}, as such the gain/loss is a function of frequency and the  gain/loss causes the real part of the dielectric constant to be dispersive. To quantify the amount of gain/loss, the gain/loss parameter used is the imaginary part of the refractive index at the atomic transitional frequency which in this case has been associated with the Bragg frequency \index{Bragg frequency} so that $n''(\omega_\sigma/(2\pi))=n''(f_B)$. The value of $n''(f_B)$ can be calculated directly by substituting $\omega\rightarrow(2\pi f_B)$ to Eq. (\ref{eq:realimagcond}) and (\ref{eq:simpdielcons}). 
	
	The performance of the PTBG is depicted in Fig. \ref{fig:06_ptbg_t_r_evolve_realistic_gl} for different values of gain/loss. Figure \ref{fig:06_ptbg_t_r_evolve_realistic_gl} considers the transmittance, $T$, and reflectance for the left, $R_L$, and right, $R_R$, incidence for increasing values of gain/loss parameter for (a) a passive structure, (b-f) $n''(f_B)=0.0045$, $0.015$, $0.02$, $0.022$ and $0.0245$. The transmittance, reflectance left and reflectance right are depicted on the top, middle and bottom panels respectively. It is noted that the transmission for both left and right incidence is the same as for a reciprocal system, \index{reciprocal system} and is shown here as $T$. In contrast to the non-dispersive PTBG structure, depicted in Fig. \ref{fig:05_ptbg_t_r_evolve}, it can be seen from the top panel of Fig. \ref{fig:06_ptbg_t_r_evolve_realistic_gl}(d) that for a dispersive PTBG system the unidirectional operation \index{unidirectional invisibility} occurs only at a single point. Moreover, the reflectance for left incidence $R_L$, increases as the gain/loss parameter increases, although in the dispersive case most amplification of $R_L$ occurs at $f_B$. A further look at the first five panels on the bottom row of Fig. \ref{fig:06_ptbg_t_r_evolve_realistic_gl}, reveals that in general $R_R$ decreases as the gain/loss parameter $n''(f_B)$ increases. In addition, Fig. \ref{fig:06_ptbg_t_r_evolve_realistic_gl}(f) shows that for $n''(f_B)=0.0245$ both transmittance and reflectance approach infinity regardless of the direction of incidence; operation at this point is related to the CPAL operation case where the spectral singularity occurs.                   
	
	\begin{figure}[th!]	
		\centering
		\begin{overpic}[width=0.6\textwidth,tics=5]{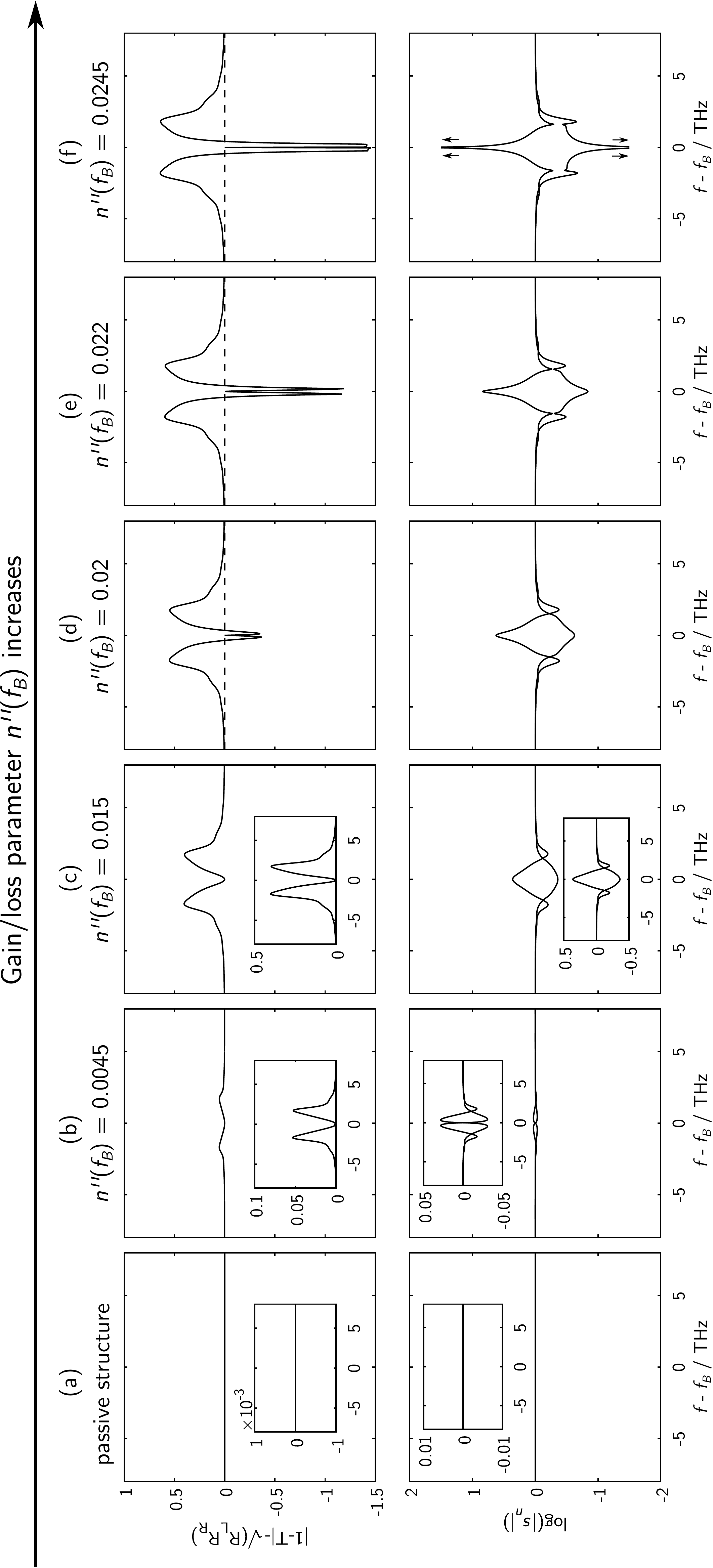}
		\end{overpic}
		\caption[The spectral behaviour of a dispersive PTBG.]{ {The spectral behaviour of a dispersive PTBG. The top panels show the difference between the left and right terms of generalised conservation relation and the bottom panels show the magnitude of the eigenvalues of the scattering matrix $\mathbf{S}$. Six different value of gain/loss parameters are considered ({a}) passive structure,  ({b}-{f}) $n''(f_B)=0.0045$, $0.015$, $0.02$, $0.022$ and $0.0245$. The insets show the magnified spectra.}}
		\label{fig:06_ptbg_eigenval_realistic_gl}
	\end{figure}
	
	To further analyse the impact of dispersion upon the spectral performance of the PTBG, Fig. \ref{fig:06_ptbg_eigenval_realistic_gl} plots the conservation relations Eq. (\ref{eq:generalconserv}) and the magnitude of the eigenvalue of the scattering matrix $\mathbf{S}$ \index{$\mathbf{S}$-matrix} for different gain/loss parameters as in Fig. \ref{fig:06_ptbg_t_r_evolve_realistic_gl}. It was discussed in Section \ref{sec:noncausalPTBG} that a \PT-symmetric scattering system has to satisfy the generalised conservation relation \index{generalised conservation relation} Eq. (\ref{eq:generalconserv}) so that the difference between the left and right hand-sides of the equation remains zero. In the dispersive PTBG system, the top panel of Fig. \ref{fig:06_ptbg_eigenval_realistic_gl} shows that the generalised conservation relation is only valid at a single frequency $f_B$.   
	
	The bottom panels of Fig. \ref{fig:06_ptbg_eigenval_realistic_gl} show the magnitude of the eigenvalues of the scattering matrix $|s_n|$ where $n=\{1,2\}$. As a reminder, it was discussed in more detail in Sections \ref{sec:pttransition} and \ref{sec:noncausalPTBG} that a \PT-symmetric scattering system may undergo a spontaneous symmetry breaking \index{spontaneous symmetry breaking} as the gain/loss parameter increases. These different symmetry phases are determined by the magnitude of the eigenvalues of the $\mathbf{S}$-matrix, so that in the symmetric phase the eigenvalues are unimodular ($|s_n|=1$) while in the broken-symmetry phase the eigenvalue is not-unimodular. In the non-dispersive PTBG structure, it was shown that the eigenvalues of the $\mathbf{S}$-matrix are unimodular until a certain value of gain/loss parameter, with operation beyond this point leading to a splitting in the value of $|s|$ which is depicted as an ``egg-shaped'' spectrum. In the dispersive PTBG system, it can be seen from the bottom panels of Fig. \ref{fig:06_ptbg_eigenval_realistic_gl} that the eigenvalues are in general not-unimodular even with a small gain/loss parameter; it can however be seen in detail from the inset of Fig. \ref{fig:06_ptbg_eigenval_realistic_gl}(b, bottom) that at a frequency $f_B$ the eigenvalues are still unimodular for $n''=0.045$. It implies that even with a small amount of gain/loss, the \PT-symmetry can occur \textit{only} at a single frequency $f_B$. From Fig. \ref{fig:06_ptbg_eigenval_realistic_gl}(f, bottom), it can be seen that there exists a spectral singularity, which is related to the CPAL point operation, that also appeared in Fig. \ref{fig:06_ptbg_t_r_evolve_realistic_gl} as the transmittance and reflectance coefficients approach infinity.           
	
	In order to understand the reason why the \PT-symmetric behaviour is only observed at a single isolated frequency in the PTBG structure with a dispersive causal gain/loss medium, recall that a \PT-symmetric structure requires a spatially modulated dielectric constant $\bar{\varepsilon}(x)=\bar{\varepsilon}'(x)+j\bar{\varepsilon}''(x)$ as, 
	\begin{align}
	\bar{\varepsilon}'(-x) &= \bar{\varepsilon}'(x) \label{eq:realdiel} \\
	\bar{\varepsilon}''(-x) &= -\bar{\varepsilon}''(x) \label{eq:imagdiel} 
	\end{align}
	That is, the real part of permittivity has to be an even function in space while the imaginary part is an odd function in space and both condition occur \textit{independently of frequency}. Moreover, one also needs to consider that the material permittivity has to satisfy the Kramers-Kronig relation\cite{Landau1984,Zyablovsky2014}, so that the modified Kramers-Kronig relations \index{Kramers-Kronig relations} is now given by:
	\begin{align}
	\varepsilon'(\omega,x) &=  \varepsilon_0 + \frac{1}{\pi} \text{ p.v.}\int_{-\infty}^{\infty} \frac{\varepsilon''(\Omega,x)}{\Omega-\omega} d\Omega  \label{eq:kkrelation1a} \\
	\varepsilon''(\omega,x) &= -\frac{1}{\pi}  \text{ p.v.}\int_{-\infty}^{\infty} \frac{\varepsilon'(\Omega,x)}{\Omega-\omega} d\Omega  
	\label{eq:kkrelation1b}
	\end{align}
	Considering operation at a real frequency $\omega$ and substituting $x\rightarrow-x$, Eq. (\ref{eq:kkrelation1a}) becomes, 
	\begin{align}
	\varepsilon'(\omega,-x) 
	&=  \varepsilon_0 + \frac{1}{\pi} \text{ p.v.}\int_{-\infty}^{\infty} \frac{\varepsilon''(\Omega,-x)}{\Omega-\omega} d\Omega   
	\label{eq:modkramers}
	\end{align}
	Further, substituting the condition Eq. (\ref{eq:imagdiel}) into Eq. (\ref{eq:modkramers}), it can be shown that 
	\begin{align}
	\varepsilon'(\omega,-x) 
	&=   \varepsilon_0 - \frac{1}{\pi} \text{ p.v.}\int_{-\infty}^{\infty} \frac{\varepsilon''(\Omega,-x)}{\Omega-\omega} d\Omega   
	\end{align}
	from which follows the condition of 
	\begin{align}
	\text{ p.v.}\int_{-\infty}^{\infty} \frac{\varepsilon''(\Omega,-x)}{\Omega-\omega} d\Omega = 0 
	\label{eq:ptcausalcond}
	\end{align}
	Equation (\ref{eq:ptcausalcond}) means that the \PT-symmetric condition Eq. (\ref{eq:realdiel}) and Eq. (\ref{eq:imagdiel}) can not be satisfied for an \textit{infinite} frequency interval except for the case of $\varepsilon'(\omega,x)=\varepsilon_0(\omega,x)$ and $\varepsilon''(\omega,x)=0$, hence a continuous medium. The conditions Eq. (\ref{eq:realdiel}) and Eq. (\ref{eq:imagdiel}) can, however, be satisfied at a single frequency associated with the resonant behaviour of the medium. 
	
	\subsection[Time-domain Modelling of a PTBG Using the TLM Method]{Time-domain Modelling of a PTBG Using the TLM Method} \label{sec:tlmptbg}
	
\begin{figure}[t]
		\begin{overpic}[width=1\textwidth,tics=5]{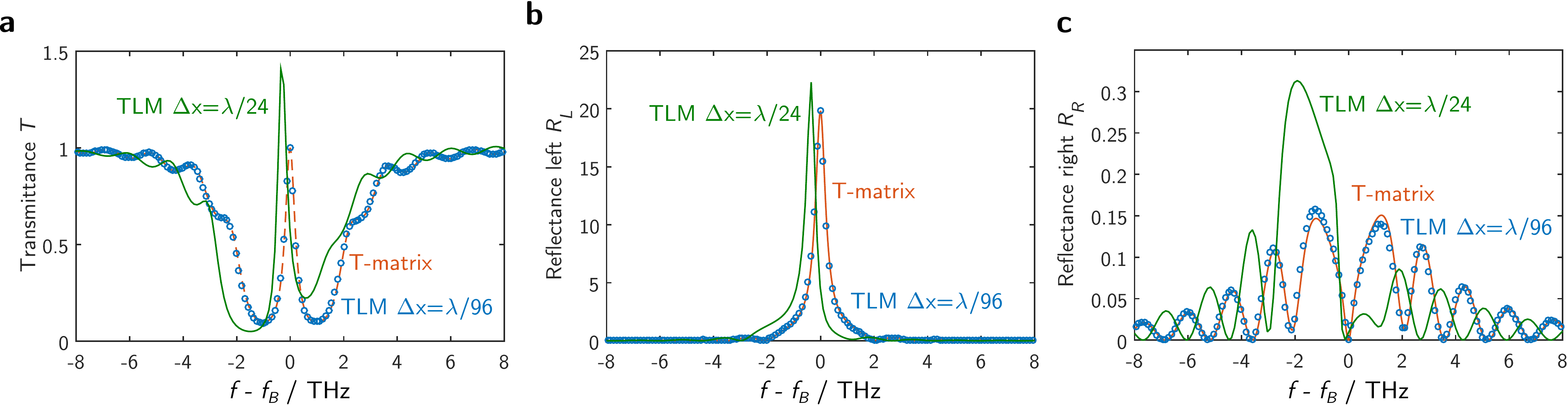}
		\end{overpic}
		\centering
		\caption[The impact of TLM mesh discretisation to the spectral response of PTBG.]{{The accuracy of the TLM to model PTBG structure. The ({a}) transmittance, ({b}) reflectance left $R_L$ and ({c}) reflectance right $R_R$ of PTBG with dispersive gain/loss. For reference results from the analytic T-matrix method is also included.}}
		\label{fig:06_tlm_ptbg_validation}
	\end{figure}

	This section demonstrates the application of the Transmission-Line Modelling (TLM) method to model the dispersive PTBG structure in the time-domain. In order to demonstrate the dependence between the accuracy of the TLM method when modelling a PTBG structure upon the discretisation parameter, the spectral performance of a PTBG under unidirectional invisibility operation is shown in Fig. \ref{fig:06_tlm_ptbg_validation}. The unidirectional operation point here refers to the operation when the real part modulation of the refractive index is equal to the gain/loss parameter. The analytical calculation using the Transfer-matrix (T-matrix) method \cite{Collin1991} at this operation point is displayed in Fig. \ref{fig:06_ptbg_t_r_evolve_realistic_gl}(d). 
	
	Figure \ref{fig:06_tlm_ptbg_validation}(a-c) shows the transmittance $T$, reflectance left $R_L$ and reflectance right $R_R$ of PTBG operating at the unidirectional invisibility point. The PTBG structure for the TLM simulation is designed and made using the material parameters described in Subsection \ref{sec:impdispptbg} and is set to operate at the unidirectional invisibility point, i.e. the peak conductivity $|\sigma_0|=211.65$ S/m is used. The TLM simulation is excited using a single Gaussian pulse function modulated at $f=f_B$ with FWHM $20$. Different mesh discretisation parameters are used, i.e.  $\Delta x=\lambda/24$ and $\lambda/96$ where $\lambda$ is the wavelength in the medium $\lambda=\lambda_0/n_b$, to demonstrate the impact of discretisation on the spectra of the scattered light. The TLM simulation is run for 9 ps, which ensures that all of the signal has passed through the structure and provides a sufficient frequency-domain resolution. The frequency domain response is obtained by Fourier transformation of the time-domain signal. For reference, results from the analytic T-matrix method are included in the figure.
	
	It can be seen from Fig. \ref{fig:06_tlm_ptbg_validation} that fine discretisation is crucial in modelling a sub-wavelength structure using a standard TLM approach\cite{Meng2013,Meng2015}. Inadequate meshing causes the spectral response to be shifted to a lower frequency compared to the analytical results, which is usually referred to as a \textit{red-shifting} error \index{red-shifting error}. Moreover, it can be seen that when $\Delta x$ is not fine enough, not only is the spectrum red-shifted, but the amplitudes are also modified significantly. The spectral response for the TLM simulation with $\Delta x = \lambda/96$ is plotted by the blue coloured circle bullet points. The spectral response for this mesh discretisation parameter agrees well with the analytical result both in frequency and in amplitude. It can be seen that modelling a PTBG using the TLM method requires a fine discretisation parameter $\Delta x$ in order to guarantee the accuracy  of the TLM simulation. For that reason, a discretisation parameter of $\Delta x=\lambda/96$ will be used in the next section to demonstrate a switching application of a PTBG by using the TLM method.      
	
	\subsection[All-Optical Switching Using \PT-Symmetric Bragg Grating]{A Temporal Optical Switch Using the \PT-Symmetric Bragg Grating} \label{sec:ptbgswit}
	This subsection investigates the transient and dynamic behaviour of a \PT-Bragg grating where the gain is suddenly introduced into parts of the system. The PTBG considered in this subsection is as studied in the previous section. The structure is excited with a continuous wave \index{continuous wave} (CW) of constant amplitude at the Bragg frequency $f=f_B$. The choice of input signal amplitude ensures that the \PT-Bragg grating operates in the linear regime, i.e. the effect of gain saturation is negligible, $\mathbb{S}=1$.   
	
	The scenario that is modelled is as follows: initially, the Bragg grating is assumed to be uniformly lossy  $n''(f_B)=-0.02$, i.e. the gain pumping is off for a 5 ps duration, under which conditions the Bragg grating has a stop band centred at the Bragg frequency $f_B$. After the 5 ps duration, the gain is introduced as might be achieved practically by turning on the gain pumping in the gain section while masking the loss sections. After another 5 ps the same temporal switching pattern is repeated. 
	
	\begin{figure}[t]
		\begin{overpic}[width=0.7\textwidth,tics=5]{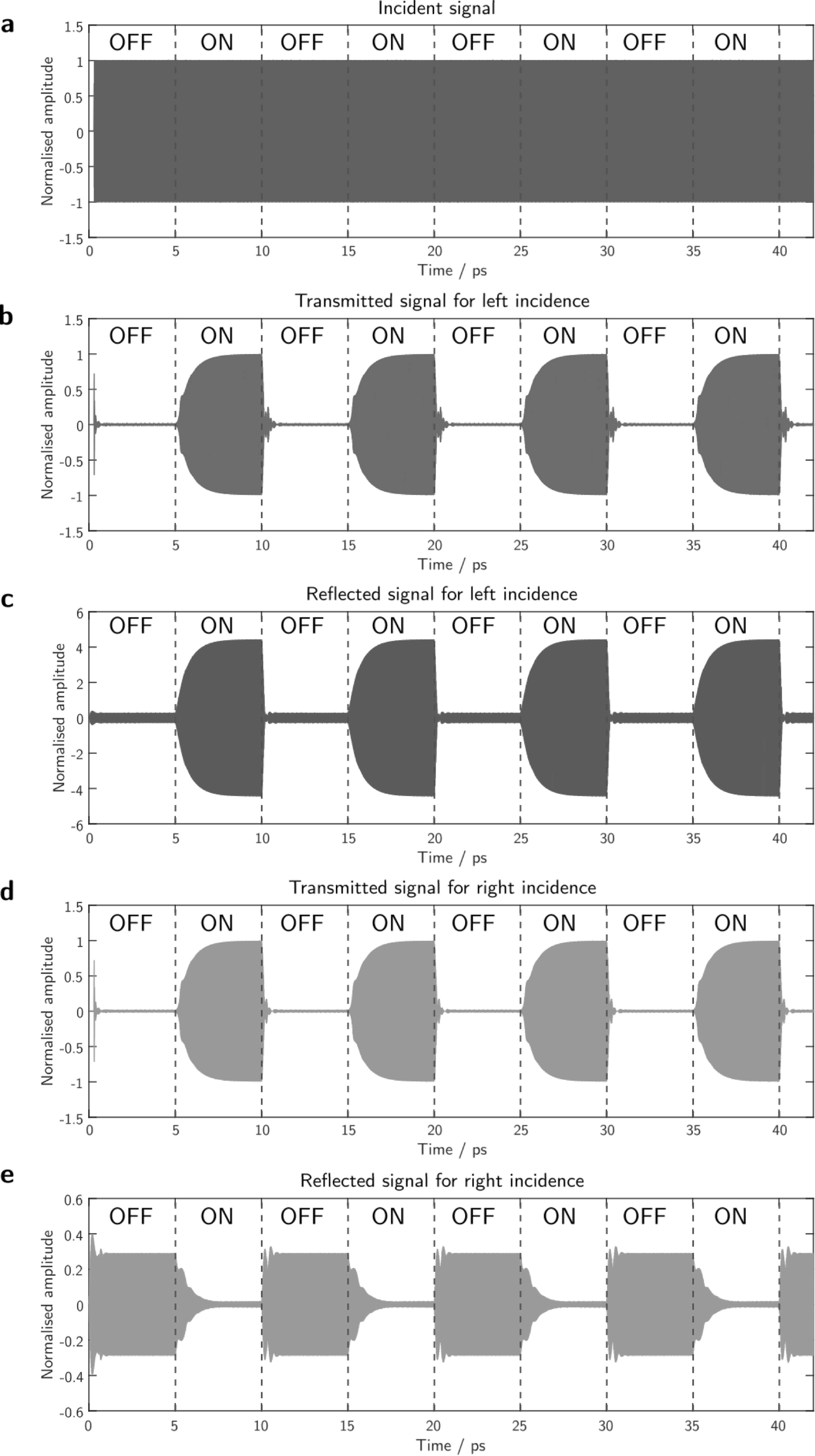}
		\end{overpic}
		\centering
		\caption[Switching application of PTBG in time-domain.]{{Switching application of PTBG in time-domain. Time-domain response of the ({a}) incident signal, ({b}) transmitted signal amplitude, ({c}) reflected signal amplitude for left incidence, ({d}) transmitted signal amplitude and ({e}) reflected signal amplitude for right incidence.}}
		\label{fig:06_tlm_switch}
	\end{figure}
	
	Figure \ref{fig:06_tlm_switch} shows the time-domain response for the input-normalised transmitted and reflected signal of the PTBG optical switch for the left and right excitations. Figure  \ref{fig:06_tlm_switch}(a) shows the input-normalised incident signal, (b and c) the transmitted and reflected signal when the grating is excited from the left and (d and e) the transmitted and reflected signal when the grating is excited from the right. It can be seen that the transmitted signal switches from nearly 0 to $\approx1$ over a transient period of less than 1 ps.  
	
	Figure \ref{fig:06_tlm_switch}(c and e) show that the reflected signal for left incidence has increased in the presence of gain but that the reflected signal for the right incident has sharply reduced to almost zero. The time-domain simulation results confirm that when the gain pumping beam is turned on, the grating transforms to a PTBG operating at the unidirectional invisibility point, and when excited from the right, its response will change from purely reflective to all transmitting and thus exhibit a switch-like behaviour. It is also emphasised that this is achieved when the grating is operated at the Bragg frequency $f_B$ with a background medium of $n_b$ and is the first demonstration of a temporal \PT-Bragg grating switch using a numerical time-domain code \cite{Phang2013}. Moreover, Fig. \ref{fig:06_tlm_switch} demonstrates that switching on gain in the grating in the real-time triggers a switch-like response from the grating.

\section[Non-Linear and Dispersive $\mathcal{PT}$-Bragg Grating]{Non-Linear and Dispersive Parity-Time Bragg Grating for Optical Signal Processing Applications}
\label{chap:nonlineardispersiveptbg}
This section considers modelling PTBG structures with strong signal excitation. As such there is the interplay between gain/loss saturation and Kerr \index{Kerr effect} non-linearity. The section starts with the implementation of a Duffing non-linear \index{Duffing non-linear} model within the time-domain TLM method. A non-linear PTBG configurations is studied which shows potential application as a memory device \index{memory device}.	
	
	\subsection{TLM Model for Non-Linear Medium}                 
	Mathematically, non-linear electromagnetic interactions with a material can be described by the polarisation of the material which behaves in a non-linear manner in the presence of strong optical electric field\cite{Joseph1997,Siegman1986,Liu2005,Iizuka2002,Saleh2007}. As such one can expand the dielectric polarisation \index{dielectric polarisation} in terms of its linear and non-linear terms as\cite{Joseph1997,Siegman1986},  
	\begin{align}
	\bm{P}_e = \underbrace{\chi_{eL}\bm{|E|}}_{\bm{P}_L} + \underbrace{\chi_e^{(2)}\bm{|E|}^2 + \chi_e^{(3)}\bm{|E|}^3 +\cdots}_{\bm{P}_{NL}}     
	\label{eq:nonlinpol} 
	\end{align}
	In Eq. (\ref{eq:nonlinpol}), $\bm{P}_L$ denotes the linear polarisation while $\bm{P}_{NL} $ represents the non-linear dielectric polarisation interaction, which may include different orders of non-linearity. Consider that the non-linear polarisation $\bm{P}_{NL}$ is modelled through the Duffing polarisation $\bm{P_D}$\cite{FatkhullaKh.Abdullaev2005,Conti2004,Janyani2004a,Janyani2005a,Janyani2004,Paul2002},    
	\begin{align}
	\frac{\partial^2 \bm{P_D}}{\partial t^2} + 2\delta \frac{\partial \bm{P_D}}{\partial t} + \omega_{0D}^2 \bm{P_D} f_D\left(\bm{P_e}\right)  = \varepsilon_0 \Delta\chi_{e0} \omega_{0D}^2\bm{E}
	\label{eq:genduff}
	\end{align}
	where $\bm{P_D}$ and $\bm{E}$ are the Duffing non-linear polarisation and electric field vector quantities which are both functions of space and time. The parameters $\omega_{0D}$ and $\delta$ are related to the Duffing polarisation angular resonant frequency and the damping constant, $\Delta\chi_{e0}$ denotes the dielectric susceptibility measured at zero frequency (DC). The function $f_D\left(\bm{P_e}\right)$ denotes the non-linear terms of the Duffing polarisation which depend on the total dielectric polarisation $\bm{P_e}$. The application of the Duffing equation to model non-linear material properties has been extensively analysed and shown to be superior to the Kerr model of a non-linear material \cite{FatkhullaKh.Abdullaev2005,Conti2004,Kazantseva2005,Janyani2004,Janyani2004a,Janyani2005a,Paul2002}. This is mainly due to the fact that the Duffing model incorporates both the non-linear and dispersive nature of the material response and thus is closer to realistic material properties \cite{Conti2004,Kazantseva2005,Janyani2004,Janyani2004a,Janyani2005a,Paul2002}.       
	
	For the case of a one-dimensional problem, with the electric field polarised in the $y$-direction, the Duffing equation (\ref{eq:genduff}) can be simplified to, 
	\begin{align}
	\frac{\partial^2 P_{Dy}}{\partial t^2} + 2\delta \frac{\partial P_{Dy}}{\partial t} + \omega_{0D}^2 P_{Dy} f_D\left(P_{ey}\right)  = \varepsilon_0 \Delta\chi_{e0} \omega_{0D}^2E_y 
	\label{eq:modduff}
	\end{align} 
	For the particular case of $f_D\left(P_{ey}\right)=1$, the Duffing polarisation is linear and dispersive with a Lorentzian \index{Lorentzian} type of dispersion and, by performing a Fourier transformation the complex dielectric permittivity can be obtained as,
	\begin{align}
	\varepsilon(\omega) =\frac{\mathcal{P}_{Dy}}{\mathcal{E}_y} = \frac{\Delta\chi_{e0}  \omega_{0D}^2}{2j\delta\omega+(\omega_{0D}^2-\omega^2)}\varepsilon_0, \quad \text{ when}\quad f_D = 1 
	\label{eq:duffloren}
	\end{align}
	In Eq. (\ref{eq:duffloren}) the field quantities $\mathcal{P}_{Dy}$ and $\mathcal{E}_y$ are the Fourier transformed Duffing polarisation $P_{Dy}$ and electric field $E_y$ which are both complex. 
	
	For the more general case of a non-linear problem, $f_D\left(P_{ey}\right)\ne1$, different non-linear functions $f_D\left(P_{ey}\right)$ have been used and analysed such as in\cite{Conti2004,Kazantseva2005,Janyani2004,Janyani2004a,Janyani2005a,Paul2002},
	\begin{align}
	\text{Exponential non-linearity } &\text{:}\quad f_D\left(P_{ey}\right)=e^{\alpha|P_{ey}|^2}  \label{eq:expnonlin}\\
	\text{Polynomial non-linearity } &\text{:}\quad f_D\left(P_{ey}\right)=1+\alpha|P_{ey}|^2  \label{eq:polnonlin}
	\end{align}     
	where, $\alpha$ denotes the Duffing non-linearity parameter, so that $\alpha=0$ defines the linear case. It is important to note that the polynomial non-linearity function Eq. (\ref{eq:polnonlin}) is an approximation of the exponential non-linear function Eq. (\ref{eq:expnonlin}). The polynomial non-linearity function Eq. (\ref{eq:polnonlin}) is only the first two terms of the Taylor expansion of the exponential non-linearity Eq. (\ref{eq:expnonlin}). This approximation is thus valid only for small values of $P_{ey}$. 
	
	The polynomial approximation of the Duffing non-linear polarisation shows an association with the Kerr-type non-linearity\cite{Janyani2005a}. To show this, consider the Kerr non-linear effect which is typically expressed as the instantaneous perturbation of the real part of the refractive index as 
	\begin{align}
	n(t)=n_L \pm n_2 I(t),
	\label{eq:kerrrefidx}
	\end{align}
	where $n_L$ denotes the constant linear refractive index which is the total of the asymptotic contribution at DC and infinity, $n_L=\sqrt{\chi_{e\infty}+\Delta \chi_{e0}+1}$. The Kerr non-linear constant $n_2$ is given in units of $\text{m}^2/$Watt. The parameter $I$ is the instantaneous intensity which is given previously by Eq. (\ref{eq:intensity1d}). The total dielectric polarisation in the presence of instantaneous Kerr non-linearity \index{Kerr non-linearity} is, 
	\begin{align}
	P_{ey} &= \varepsilon_0(n^2(t)-1)E_y 
	\end{align}
	which can be approximated by,
	\begin{align}
	P_{ey} \approx \varepsilon_0(n^2_L-1)E_y + 2\varepsilon_0n_Ln_2IE_y, \quad \text{as} \quad (n_2I)^2 \rightarrow 0  
	\end{align}
	Expanding the linear refractive index, the dielectric polarisation can be expressed as, 
	\begin{align}
	P_{ey} =\varepsilon_0\left(\chi_{e\infty}+\Delta \chi_{e0}\right)E_y + 2\varepsilon_0n_Ln_2IE_y \label{eq:kerrpol}
	\end{align}
	By substituting the intensity $I$ defined in Eq. (\ref{eq:intensity1d}) into Eq. (\ref{eq:kerrpol}) and comparing it to Eq. (\ref{eq:nonlinpol}), the dielectric susceptibility can be obtained as,
	\begin{align}
	\chi_{eL} &= (\chi_{e\infty}+\Delta\chi_{e0})\varepsilon_0  \\
	\chi_e^{(3)} &= \frac{\left(\chi_{e\infty}+\Delta \chi_{e0}+1\right)n_2}{\eta_0}\varepsilon_0 
	\end{align}
	where, $\chi_e^{(3)}$ denotes the Kerr non-linear susceptibility constant. 
	
	In order to find the association between the Duffing non-linearity \index{Duffing non-linear} and the Kerr non-linearity, consider a dispersion-less non-linear Duffing polarisation by substituting $\partial P_{Dy}/\partial t\rightarrow0$ into Eq. (\ref{eq:modduff}). The dielectric polarisation can be obtained as, 
	\begin{align}
	f_D(P_{ey}) P_D = \varepsilon_0 \Delta \chi_{e0} E_y
	\label{eq:linduff}
	\end{align}
	By substituting the polynomial approximation for non-linear polarisation Eq. (\ref{eq:polnonlin}) into Eq. (\ref{eq:linduff}) it can be shown that, 
	\begin{align}
	P_{ey} +\alpha P_{ey}^2 ( P_{ey} - P_\infty )  = \varepsilon_0 (\chi_{e\infty}+\Delta\chi_{e0}) E_y 
	\label{eq:duffpol}
	\end{align} 
	where $P_\infty=\varepsilon_0 \chi_{e\infty} E_y$ is the asymptotic polarisation contribution at infinity. By direct comparison of Eq. (\ref{eq:duffpol}) and Eq. (\ref{eq:kerrpol}), the relation between the Duffing non-linear parameter $\alpha$ and the Kerr non-linear parameter $n_2$ at the small-signal excitation can be found as, 
	\begin{align}
	\alpha = -\frac{n_L^2 n_2}{\varepsilon_0^2 \eta_0 (\chi_{e\infty}+\Delta\chi_{e0})^2  \Delta\chi_{e0}} \quad \text{where, } \quad 
	n_L^2=\chi_{e\infty}+\Delta \chi_{e0}+1 
	\label{eq:duffalpha}
	\end{align}  
	
	In order to implement the Dufffing model within the TLM method, a digital filter \index{digital filter} representing Duffing non-linear polarisation model Eq. (\ref{eq:modduff}) is now developed.  
	The normalised Duffing model is given by (see Subsection \ref{sec:realisticgain}), 
	\begin{align}
	\frac{\partial^2 p_{Dy}}{\partial T^2} + K_{D1} \frac{\partial p_{Dy}}{\partial T} + K_{D2} p_{Dy} f_D\left(p_{ey}\right)  = K_{D3} V_y
	\label{eq:modduffnorm}
	\end{align}
	where, $p_{Dy}=-\frac{P_{Dy}\Delta x}{\varepsilon_0}$ is the normalised Duffing polarisation. The dimensionless constants in Eq. (\ref{eq:modduffnorm}) are defined as, 
	\begin{align*}
	K_{D1} = 2\delta\Delta t, \quad
	K_{D2} = \left(\omega_{0D} \Delta t\right)^2, \quad
	K_{D3} = \Delta\chi_{e0} \left(\omega_{0D} \Delta t\right)^2 
	\end{align*}  
	where $\Delta t $ is the TLM time-step. By an application of the bilinear $\mathcal{Z}$-transform \index{$\mathcal{Z}$-transform} Eq. (\ref{eq:normz}) and after some re-arrangement, the Duffing model in the $\mathcal{Z}$-domain is given by, 
	\begin{align}
	p_{Dy}K_{D4} + p_{Dy}K_{D2} f_D\left(p_{ey}\right) + z^{-1}S_{D1}= K_{D3}V_y,
	\label{eq:duffdig}
	\end{align}
	where, 
	\begin{align}
	K_{D4} &= \left(4  +2K_{D1} \right)  \\
	S_{D1} &= \left[p_{Dy}\left(-8  + 2K_{D2}f_D\right)-2K_{D3}V_y\right]+ z^{-1} S_{D2}  \\
	S_{D2} &= \left[p_{Dy}\left(4 -2K_{D1}+K_{D2} f_D\right)-K_{D3}V_y\right] 
	\end{align}
	The normalised non-linear Duffing exponential function is given by, 
	\begin{align}
	f_D\left(p_{ey}\right)=e^{\alpha|p_{ey}|^2} \label{eq:expnonlinnorm}
	\end{align}
	with the Duffing non-linear parameter $\alpha$ as given in Eq. (\ref{eq:duffalpha}). Equation (\ref{eq:duffdig}) is the transcendental non-linear polarisation equation which will be solved simultaneously with the TLM main equation by an iterative method. 
	
	Upon substituting the linear and non-linear dielectric polarisation into the the 1D-TLM scattering equation (\ref{eq:1dgainmod1}), and after some algebraic arrangement, Eq. (\ref{eq:1dgainmod1}) can be expressed as 
	\begin{align}
	2\left(V_y^r - V_y\right)\left(1+z^{-1}\right) = \left(1+z^{-1}\right) g_eV_y +2 \left(1-z^{-1}\right) \left(\chi_{e\infty}V_y +p_{Dy}\right)
	\label{eq:1dgainmod3}
	\end{align}  
	By substituting the dispersive conductivity model Eq. (\ref{eq:ge3}), Eq. (\ref{eq:1dgainmod3}) becomes 
	\begin{equation}
	\begin{split}
	2\left(V_y^r - V_y\right)\left(1+z^{-1}\right) =&\\  \left\lbrace g_{e0} + z^{-1}(g_{e1} + \bar{g}_e(z)) \right\rbrace V_y +2 \left(1-z^{-1}\right) \left(\chi_{e\infty}V_y +p_{Dy}\right) &
	\label{eq:1dgainmod4}
	\end{split}
	\end{equation}
	By grouping the present and past variables in Eq. (\ref{eq:1dgainmod4}), it can be shown that 
	\begin{align}
	2V_y^r + z^{-1}\left(2V_y^r  + K_{e1}V_y  - \bar{g}_e(z) V_y + 2p_{Dy}\right) = K_{e2} V_y  +2p_{Dy} 
	\label{eq:1dgainmod5}
	\end{align}
	where the constants are defined as, 
	\begin{align}
	K_{e1} &= -(2+ g_{e1} -2\chi_{e\infty})  \\
	K_{e2} &= 2+ g_{e0}+2\chi_{e\infty} 
	\end{align}
	By further calling the sum of all the past variables in Eq. (\ref{eq:1dgainmod5}) as, 
	\begin{align}
	S_{ey} & = 2V_y^r  + K_{e1}V_y  +S_{ec} + 2p_{Dy} \\
	S_{ec} & = - \bar{g}_e(z) V_y 
	\end{align}
	equation (\ref{eq:1dgainmod5}) can be simplified further as, 
	\begin{align}
	K_{e2} V_y  +2p_{Dy}  = 2V_y^r + z^{-1}S_{ey}
	\label{eq:1dgainmod6}
	\end{align}
	The equations (\ref{eq:1dgainmod6}) and Eq. (\ref{eq:duffdig}) are two coupled equations with two unknown variables $V_y$ and $p_{Dy}$, 
	\begin{align}
	\begin{cases}
	K_{e2} V_y +2p_{Dy}  = 2V_y^r + z^{-1}S_{ey}  \\
	K_{D3}V_y = p_{Dy}K_{D4} + p_{Dy}K_{D2} f_D + z^{-1}S_{D1}
	\end{cases}
	\end{align}
	which are now ready to be solved simultaneously for $p_{Dy}$ by an iterative method, e.g. either the Newton-Rhapson or Bi-section methods \cite{Press2002}. The nodal voltage $V_y$ can be subsequently obtained by substituting the solved $p_{Dy}$ back into Eq. (\ref{eq:1dgainmod6}). In this present work, a combined Newton-Rhapson and Bi-section method is used, capitalising on the fast convergence of the Newton-Rhapson method and the stability of the Bi-section method; for detail on the implementation of the method, readers are referred to \cite{Press2002}.
	
	\subsection[Non-Linear \PT-Bragg Grating]{Non-Linear \PT-Bragg Grating For a Memory Device} 	
	Consider now a non-linear version of the \PT-Bragg grating (NPTBG); the NPTBG is similar to that studied in Subsection \ref{sec:impdispptbg} but a uniform Kerr non-linearity \index{Kerr non-linearity} is now added throughout the unit cell. We will use the Duffing \index{Duffing non-linear} model to model the Kerr non-linearity of the medium. The refractive index in a single period, $n_G$, along the propagation direction $x$ is now modified as, 
	\begin{align}
	n_G(x,\omega,I,t)
	\begin{cases}
	n_{\text{hi}}(\omega) + n_2 I(x,t) - j\dfrac{c_0}{\omega}\alpha(\omega,I), &\quad x<\dfrac{\Lambda}{4} \\
	n_{\text{lo}}(\omega) + n_2 I(x,t) - j\dfrac{c_0}{\omega}\alpha(\omega,I), &\quad 
	\dfrac{\Lambda}{4}<x<\dfrac{\Lambda}{2}\\
	n_{\text{lo}}(\omega) + n_2 I(x,t) +j\dfrac{c_0}{\omega}\alpha(\omega,I), &\quad \dfrac{\Lambda}{2}<x<\dfrac{3\Lambda}{4}\\
	n_{\text{hi}}(\omega) + n_2 I(x,t) +j\dfrac{c_0}{\omega}\alpha(\omega,I), &\quad \dfrac{3\Lambda}{4}<x<\Lambda
	\end{cases} 
	\label{eq:ptbgrefind}
	\end{align}  
	where $n_{\text{hi}}$ and $n_{\text{lo}}$ are frequency dependent complex high and low refractive indices, whose parameters are summarised in Table \ref{tab:matparams}, $n_2$ is the Kerr non-linearity coefficient, $I$ is the input signal intensity and $\pm\alpha$ denotes the gain (-) and loss (+) in the grating lattices. The gain/loss is modelled using the realistic gain/loss conductivity model described in Subsection \ref{sec:realisticgain} so using the relation Eq. (\ref{eq:gain}) between the gain/loss $\alpha$ and the dispersive imaginary part of the refractive index, the peak value of gain/loss parameter $\alpha_0$ can be defined as
	\begin{align}
	\alpha_0=\dfrac{\omega_\sigma}{c_0} n''(\omega_\sigma)  
	\label{eq:gainparam}
	\end{align}
	\begin{table}[bp!]\centering
		\caption{{Material parameters used to model non-linear Bragg grating.}}
		\begin{tabular}{@{}lcc@{}} 
			\toprule
			Parameters            		& Low refractive index  & High refractive index  \\ 
			\midrule
			$\chi_{e\infty}$    		& $2.5$            & $2.8$ \\
			$\Delta\chi_{e0}$   		& $7.5$            & $7.5$ \\
			$\delta$ (rad/ps)           & $0.0923$         & $0.0923$  \\
			$\omega_{0D}$ (rad/ps)      & $4614.4$ 	       & $4614.4$  \\	
			\bottomrule
		\end{tabular}
		\label{tab:matparams}
	\end{table}
	
	For modelling purposes, the gain/loss material parameters are set as follows: the  atomic transition frequency of the gain/loss material is set to coincide with the Bragg frequency \index{Bragg frequency}, i.e. $\omega_{\sigma}=2\pi f_B$ with the time relaxation constant $\tau = 0.1$ ps similar to that used in \cite{Hagness1996,Phang2014d} and the saturation intensity is set at $I_\text{sat}=5\times 10^{13} \text{ W/m}^2$.
	
	\begin{figure}[tp]
		\begin{overpic}[width=0.9\textwidth,tics=5]{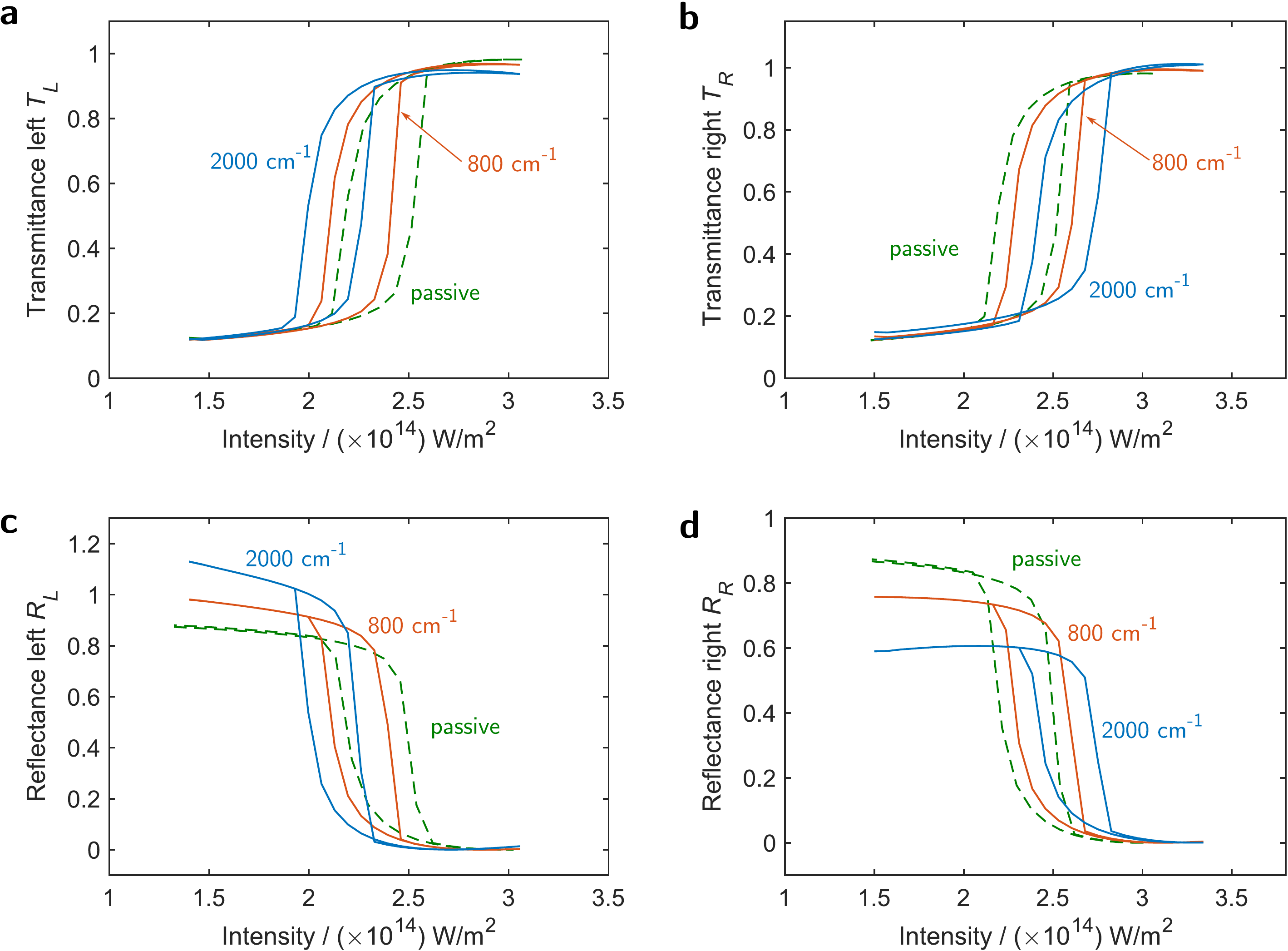}
		\end{overpic}
		\centering
		\caption[Hysteresis of non-linear PTBG with high saturation intensity gain/loss material.]{{Hysteresis of non-linear PTBG with high saturation intensity gain/loss material. For the passive case and gain/loss parameter of $800\text{ cm}^{-1}$ and $2000\text{ cm}^{-1}$. ({a}) Transmittance $T_L$, ({c}) reflectance $R_L$, for the light incident from the left, ({b}) transmittance $T_R$, ({d}) reflectance $R_R$ for the light incident from the right of the grating. Saturation intensity is $I_\text{sat}=5\times 10^{13} \text{ W/m}^2$. Dashed line represents the response of the equivalent passive NBG for reference.} }
		\label{fig:07_hysteresis_ptbg_sat_5e13}
	\end{figure}
	
	Figure \ref{fig:07_hysteresis_ptbg_sat_5e13} shows (a) transmittance $T_L$ and (c) reflectance $R_L$ for the left incidence case and (b) transmittance $T_R$ and (d) reflectance $R_R$ for the right incident signal as a function of input signal intensity and for different gain and loss parameter $\alpha_0$. For comparison, the response of a \textit{passive} non-linear Bragg grating (NBG) (i.e. one without gain and loss, $\alpha_0=0$ ) is depicted by dashed lines. 
	
	In order to obtain bistable operation the input signal frequency is set to be at the right flank of the band-gap \cite{Suryanto2003,Phang2014d}, in which a continuous-wave (CW) operating at $f_\text{op}=337.57\text{ THz}$ is chosen. The hysteresis \index{hysteresis} is obtained by gradually increasing and decreasing the input signal intensity in a single computation. This is repeated for different gain/loss parameters, namely $\alpha_0=800\text{ cm}^{-1}$ and $2000\text{ cm}^{-1}$. 
	
	Figure \ref{fig:07_hysteresis_ptbg_sat_5e13}(a-d) shows that the NPTBG is bistable for both transmittance and reflectance regardless of the side of incidence (left or right). Figure \ref{fig:07_hysteresis_ptbg_sat_5e13}(a,c) shows that compared to a non-linear Bragg grating (NBG), the bistability \index{bistability} of the NPTBG occurs at lower input intensities for the signals incident from the left of the grating and at higher intensity for signals incident from the right side of the grating. It is noted that the transmittances for the left and right incidence are different, $T_L\ne T_R$, as shown in Fig. \ref{fig:07_hysteresis_ptbg_sat_5e13}(a,b), showing that the NPTBG does not satisfy Lorentzian reciprocity. This is due to the fact that the scattering matrix is no longer a complex-symmetric matrix \index{complex-symmetric matrix}, $\mathbf{S} \ne \mathbf{S}^T$. Furthermore, it is observed that at high intensity, both $R_L$ and $R_R$ are very low while transmittances are almost unity, implying the behaviour of a  bidirectionally transparent material at high intensity (Fig. \ref{fig:07_hysteresis_ptbg_sat_5e13}(c,d)).
	
	For an application of the NPTBG as a memory device \index{memory device}, consider a gain/loss material with a high saturation intensity as in the case studied in Fig. \ref{fig:07_hysteresis_ptbg_sat_5e13}. The NPTBG is operated at $f_\text{op}=337.57\text{ THz}$ with gain/loss parameter $\alpha_0=2000\text{ cm}^{-1}$ and with a saturation intensity $I_\text{sat}=5\times 10^{13}\text{ W/m}^2$ as in Fig. \ref{fig:07_hysteresis_ptbg_sat_5e13}. The TLM modelling was undertaken as follows: a CW light signal was excited from the left side of the NPTBG at $f_\text{op}$, and the intensity of the CW was varied throughout the simulation to emulate memory \textit{reading}, \textit{writing} and \textit{resetting} operations of the RAM device. The reading operation is set to be at $I_\text{read}=2.2\times 10^{14}\text{ W/m}^2$, the memory writing operation occurs by increasing the input intensity to $I_\text{write}=2.725\times 10^{14}\text{ W/m}^2$ while the resetting operation is achieved by decreasing the input intensity to $I_\text{reset}=1.5\times 10^{14}\text{ W/m}^2$. During the simulation each process happens for a duration of $10\text{ ps}$ and is patterned as read, write, read and reset; the same pattern is then repeated. 
	
	\begin{figure}[tb]
		\begin{overpic}[width=0.85\textwidth,tics=5]{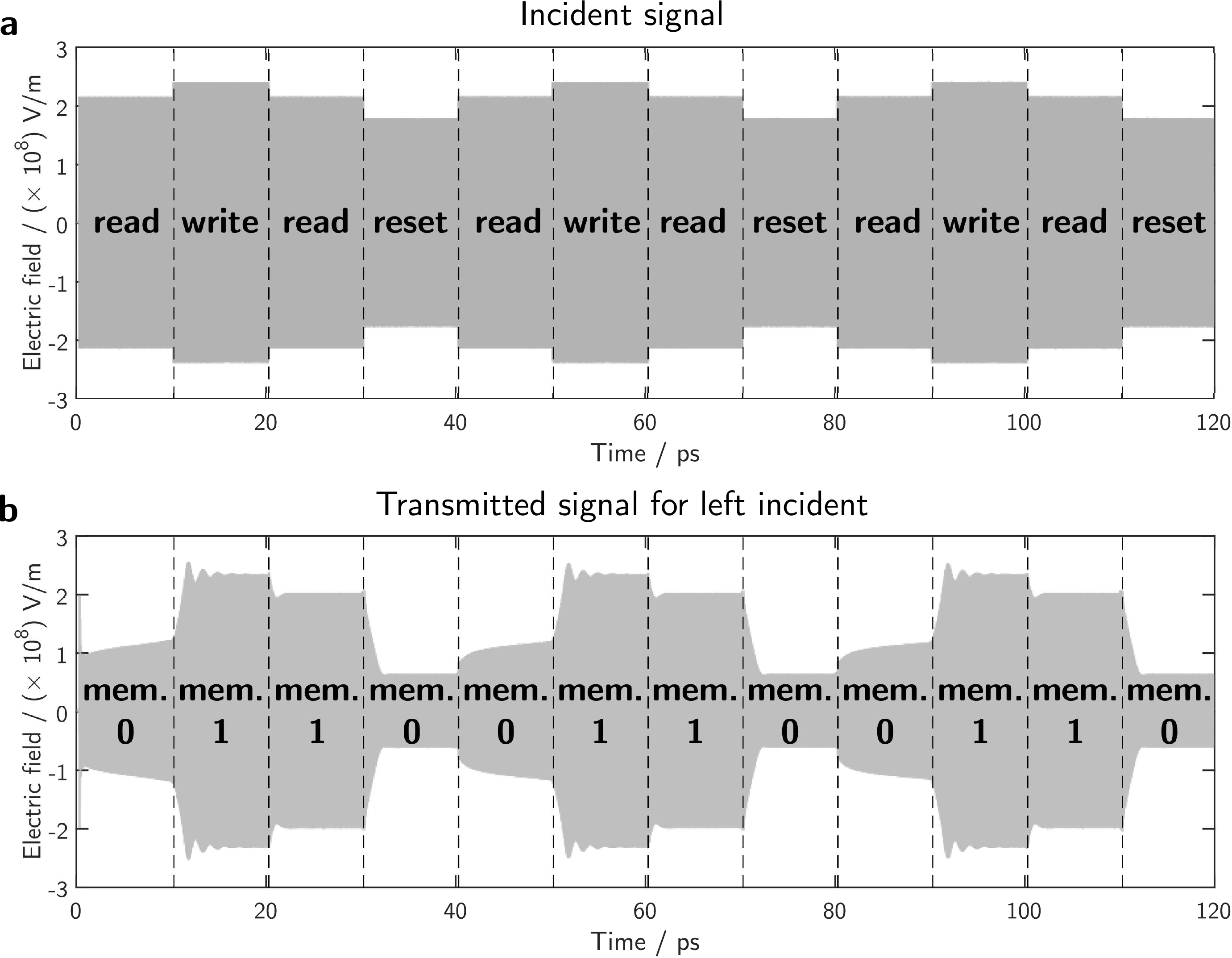}
		\end{overpic}
		\centering
		\caption[Demonstration of application of NPTBG as a memory device.]{{Demonstration of application of NPTBG as a memory device. The electric field of ({a}) the input signal and ({b}) the transmitted signal.} }
		\label{fig:07_opticalmemory}
	\end{figure}
	
	Figure \ref{fig:07_opticalmemory}(a) depicts the input signal (electric field) as a function of time, each process is labelled within the figure as \textit{read}, \textit{write} and \textit{reset} over a total simulation time of $120\text{ ps}$. The transmitted electric field is plotted in Fig. \ref{fig:07_opticalmemory}(b). It can be seen in this figure that initially the memory is in the ``0'' \textit{null} state. At $t = 10\text{ ps}$ a ``write" operation occurs by increasing the input signal intensity to achieve the ``on" state in the hysteresis, to fill the memory storage, denoted by memory ``1''. After the writing the information, the input signal intensity is reduced to the reading intensity level. It can be seen from Fig. \ref{fig:07_opticalmemory}(b) that the transmitted signal during the reading process with memory ``1'' ($20 <t<30\text{ ps}$) is higher when compared to when the memory is null ``0''($0 <t<10\text{ ps}$). By sending the reset signal (reducing the input signal intensity), the hysterisis is now at the ``off" state; as such the output during the reading process gives a small transmitted signal (memory value reset back to null). Furthermore, Fig. \ref{fig:07_opticalmemory}(b) shows that the write, read and reset operation can be performed many times with reproducible response. Figure \ref{fig:07_hysteresis_ptbg_sat_5e13} shows that by using a NPTBG grating the memory operation is performed at a lower input power compared to that would be resulted for passive non-linear Bragg grating. 
	
	Another non-linear \PT-Bragg grating structure for a logic-gate device has also been proposed in \cite{Phang2015}. The non-linear \PT-Bragg grating studied in \cite{Phang2015} was inspired by the grating structure studied by Sargent and Brozozowski \cite{Brzozowski2000}, where the grating had alternating layers of negative and positive Kerr non-linearity, but without the inclusion of gain/loss.
	
\section{Concluding remarks}
In this chapter the isomorphism between the Helmholtz and Schr\"{o}dinger equations has been reviewed. This indicates how, building on Quantum Mechanical concepts, a wide range of structures with unique and extremely interesting properties can be realised in photonic systems incorporating balanced gain and loss profiles. The scattering and transfer matrix analysis of the illustrative example of an ideal Bragg grating with a \PT-symmetric refractive index modulation gives detailed insight to the conditions under which simultaneous coherent perfect absorber-lasing operation could be obtained. In reality material properties are non-ideal. The effect of this can be studied with realistic gain/loss models implemented within a numerical Transmission-Line Modelling (TLM) method; in the non-ideal case \PT-symmetry can only occur at a single frequency. The TLM method is readily extended to include the additional effect of non-linearity and the promising use of a non-linear \PT-Bragg grating as a memory device was explored. It is inferred that photonic structures with thought-provoking functional behaviour exist even when \PT-symmetry condition cannot be met exactly.  
\label{sec:conc}

\bibliographystyle{spphys_rev}
\bibliography{references}

\begin{thebibliography}{10}
\providecommand{\url}[1]{{#1}}
\providecommand{\urlprefix}{URL }
\expandafter\ifx\csname urlstyle\endcsname\relax
  \providecommand{\doi}[1]{DOI \discretionary{}{}{}#1}\else
  \providecommand{\doi}{DOI \discretionary{}{}{}\begingroup
  \urlstyle{rm}\Url}\fi

\bibitem{Bender1998}
Bender CM, Boettcher S. {Real spectra in non-Hermitian Hamiltonians having PT
  symmetry}. \emph{Phys. Rev. Lett.} 80(24), 5243,  1998

\bibitem{Bender2002}
Bender CM, Brody DC, Jones HF. {Complex extension of quantum mechanics}.
  \emph{Phys. Rev. Lett.} 89(27), 270401,  2002

\bibitem{wang2010}
Wang QH, Chia SZ, Zhang JH. {PT symmetry as a generalization of Hermiticity}.
  \emph{J. Phys. A Math. Theor.} 43(29),  2010

\bibitem{Dizdarevic2015}
Dizdarevic D, Dast D, Haag D, Main J, Cartarius H, Wunner G. {Cusp bifurcation
  in the eigenvalue spectrum of PT-symmetric Bose-Einstein condensates}.
  \emph{Phys. Rev. A} 91(3), 033636,  2015

\bibitem{Gutohrlein2015}
Gut{\"{o}}hrlein R, Schnabel J, Iskandarov I, Cartarius H, Main J, Wunner G.
  {Realizing PT-symmetric BEC subsystems in closed Hermitian systems}. \emph{J.
  Phys. A Math. Theor.} 48(33), 335302,  2015

\bibitem{Single2014}
Single F, Cartarius H, Wunner G, Main J. {Coupling approach for the realization
  of a PT-symmetric potential for a Bose-Einstein condensate in a double well}.
  \emph{Phys. Rev. A} 90(4), 042123,  2014

\bibitem{Bagarello2013}
Bagarello F, Pantano G. {Pseudo-Fermions in an electronic loss-gain circuit}.
  \emph{Int. J. Theor. Phys.} 52(12), 4507,  2013

\bibitem{Schindler2012}
Schindler J, Lin Z, Lee JM, Ramezani H, Ellis FM, Kottos T. {PT-symmetric
  electronics}. \emph{J. Phys. A Math. Theor.} 45(44), 444029,  2012

\bibitem{Schindler2011}
Schindler J, Li~A, Zheng MC, Ellis FM, Kottos T. {Experimental study of active
  LRC circuits with PT symmetries}. \emph{Phys. Rev. A} 84(4), 040101,  2011

\bibitem{Bender2013}
Bender CM, Berntson BK, Parker D, Samuel E. {Observation of PT phase transition
  in a simple mechanical system}. \emph{Am. J. Phys.} 81(3), 173,  2013

\bibitem{Zhu2014}
Zhu X, Ramezani H, Shi C, Zhu J, Zhang X. {PT-Symmetric acoustics}. \emph{Phys.
  Rev. X} 4(3), 031042,  2014

\bibitem{Fleury2015}
Fleury R, Sounas D, Al{\`{u}} A. {An invisible acoustic sensor based on
  parity-time symmetry}. \emph{Nat. Commun.} 6, 5905,  2015

\bibitem{Poli2015}
Poli C, Bellec M, Kuhl U, Mortessagne F, Schomerus H. {Selective enhancement of
  topologically induced interface states in a dielectric resonator chain}.
  \emph{Nat. Commun.} 6, 6710,  2015

\bibitem{Bittner2012}
Bittner S, Dietz B, G{\"{u}}nther U, Harney HL, Miski-Oglu M, Richter A,
  Sch{\"{a}}fer F. {PT symmetry and spontaneous symmetry breaking in a
  microwave billiard}. \emph{Phys. Rev. Lett.} 108(2), 024101,  2012

\bibitem{Jones2012}
Jones HF. {Analytic results for a PT -symmetric optical structure}. \emph{J.
  Phys. A Math. Theor.} 45(13), 135306,  2012

\bibitem{Ramezani2010}
Ramezani H, Kottos T, El-Ganainy R, Christodoulides DN. {Unidirectional
  nonlinear PT-symmetric optical structures}. \emph{Phys. Rev. A} 82(4),
  043803,  2010

\bibitem{Lin2011}
Lin Z, Ramezani H, Eichelkraut T, Kottos T, Cao H, Christodoulides DN.
  {Unidirectional invisibility induced by PT-symmetric periodic structures}.
  \emph{Phys. Rev. Lett.} 106(21), 213901,  2011

\bibitem{Kulishov2013}
Kulishov M, Kress B, Slav{\'{\i}}k R. {Resonant cavities based on
  Parity-Time-symmetric diffractive gratings}. \emph{Opt. Express} 21(8), 68,
  2013

\bibitem{Phang2013}
Phang S, Vukovic A, Susanto H, Benson TM, Sewell P. {Ultrafast optical
  switching using parity-time symmetric Bragg gratings}. \emph{J. Opt. Soc. Am.
  B} 30(11), 2984,  2013

\bibitem{Phang2014d}
Phang S, Vukovic A, Susanto H, Benson TM, Sewell P. {Impact of dispersive and
  saturable gain/loss on bistability of nonlinear parity-time Bragg gratings}.
  \emph{Opt. Lett.} 39(9), 2603,  2014

\bibitem{Phang2015}
Phang S, Vukovic A, Benson TM, Susanto H, Sewell P. {A versatile all-optical
  parity-time signal processing device using a Bragg grating induced using
  positive and negative Kerr-nonlinearity}. \emph{Opt. Quantum Electron.}
  47(1), 37,  2015

\bibitem{Huang2013}
Huang CY, Zhang R, Han JL, Zheng J, Xu~JQ. {Type-II perfect absorption and
  amplification modes with controllable bandwidth in combined PT-symmetric and
  conventional Bragg-grating structures}. \emph{Phys. Rev. A} 89(2), 023842,
  2014

\bibitem{Rivolta2015}
Rivolta NXA, Maes B. {Diffractive switching by interference in a tailored
  PT-symmetric grating}. \emph{J. Opt. Soc. Am. B} 32(7), 1330,  2015

\bibitem{Regensburger2013}
Regensburger A, Miri MA, Bersch C, N{\"{a}}ger J, Onishchukov G,
  Christodoulides DN, Peschel U. {Observation of defect states in PT-symmetric
  optical lattices}. \emph{Phys. Rev. Lett.} 110(22), 223902,  2013

\bibitem{Longhi2011}
Longhi S. {Invisibility in PT-symmetric complex crystals}. \emph{J. Phys. A
  Math. Theor.} 44(48), 485302,  2011

\bibitem{Makris2008}
Makris KG, El-Ganainy R, Christodoulides DN. {Beam dynamics in PT symmetric
  optical lattices}. \emph{Phys. Rev. Lett.} 100(10), 103904,  2008

\bibitem{Nolting1996}
Nolting H, Sztefka G, {\v{C}}tyrok{\'{y}} J, in \emph{Integr. Photonics Res.}
  (OSA, Boston, Massachusetts, 1996) 4930 pp. 76--80

\bibitem{Ruschhaupt2005}
Ruschhaupt A, Delgado F, Muga JG. {Physical realization of -symmetric potential
  scattering in a planar slab waveguide}. \emph{J. Phys. A. Math. Gen.} 38(9),
  L171,  2005

\bibitem{Greenberg2005}
Greenberg M, Orenstein M. {Optical unidirectional devices by complex spatial
  single sideband perturbation}. \emph{IEEE J. Quantum Electron.} 41(7), 1013,
  2005

\bibitem{sukhorukov2010}
Sukhorukov AA, Xu~Z, Kivshar YS. {Nonlinear suppression of time reversals in
  PT-symmetric optical couplers}. \emph{Phys. Rev. A} 82(4), 043818,  2010

\bibitem{Nazari2011}
Nazari F, Nazari M, Moravvej-Farshi MK. {A 2x2 spatial optical switch based on
  PT-symmetry}. \emph{Opt. Lett.} 36(22), 4368,  2011

\bibitem{Kuzmiak2010}
{\v{C}}tyrok{\'{y}} J, Kuzmiak V, Eyderman S. {Waveguide structures with
  antisymmetric gain/loss profile}. \emph{Opt. Express} 18(21), 21585,  2010

\bibitem{Lupu2013}
Lupu A, Benisty H, Degiron A. {Switching using PT symmetry in plasmonic
  systems: positive role of the losses}. \emph{Opt. Express} 21(18), 192,  2013

\bibitem{Benisty2011}
Benisty H, Degiron A, Lupu A, Lustrac AD, Forget S, Besbes M, Barbillon G,
  Bruyant A, Blaize S, L{\'{e}}rondel G. {Implementation of PT symmetric
  devices using plasmonics : Principle and applications}. \emph{Opt. Express}
  19(19), 3567,  2011

\bibitem{Alaeian2014}
Alaeian H, Dionne Ja. {Parity-time-symmetric plasmonic metamaterials}.
  \emph{Phys. Rev. A} 89(3), 033829,  2014

\bibitem{Baum2015}
Baum B, Alaeian H, Dionne J. {A parity-time symmetric coherent plasmonic
  absorber-amplifier}. \emph{J. Appl. Phys.} 117(063106), 063106,  2015

\bibitem{Feng2014}
Feng L, Wong ZJ, Wang Y, Zhang X, Ma~RMRM, Wang Y, Zhang X, Ma~RMRM, Wang Y,
  Zhang X. {Single-mode laser by parity-time symmetry breaking}. \emph{Science}
  346(6212), 972,  2014

\bibitem{Chang2014}
Chang L, Jiang X, Hua S, Yang C, Wen J, Jiang L, Li~G, Wang G, Xiao M.
  {Parity-time symmetry and variable optical isolation in
  active-passive-coupled microresonators}. \emph{Nat. Photonics} 8(7), 524,
  2014

\bibitem{Longhi2014}
Longhi S, Feng L. {PT-symmetric microring laser-absorber}. \emph{Opt. Lett.}
  39(17), 5026,  2014

\bibitem{Peng2014}
Peng B, Ozdemir K, Rotter S, Yilmaz H, Liertzer M, Monifi F, Bender CM, Nori F,
  Yang L. {Loss-induced suppression and revival of lasing.} \emph{Science}
  346(6207), 328,  2014

\bibitem{phang2015b}
Phang S, Vukovic A, Creagh SC, Benson TM, Sewell PD, Gradoni G. {Parity-time
  symmetric coupled microresonators with a dispersive gain/loss}. \emph{Opt.
  Express} 23(9), 11493,  2015

\bibitem{Peng2014d}
Peng B, {\"{O}}zdemir ÅK, Lei F, Monifi F, Gianfreda M, Long GL, Fan S, Nori F,
  Bender CM, Yang L. {Parity-time-symmetric whispering-gallery microcavities}.
  \emph{Nat. Phys.} 10(5), 394,  2014

\bibitem{Phang2015d}
Phang S, Vukovic A, Creagh SC, Sewell PD, Gradoni G, Benson TM. {Localized
  single frequency lasing states in a finite parity-time symmetric resonator
  chain}. \emph{Scientific reports} 6(20499), (20499) 1,  2016

\bibitem{Zettili2009}
Zettili N, \emph{{Quantum Mechanics: Concepts and Applications}} 2nd edn. (John
  Wiley, New York, NY, 2009)

\bibitem{Yariv1989}
Yariv A, \emph{{Quantum Electronics}} 3rd edn. (John Wiley, New York, NY, 1989)

\bibitem{bender13a}
Bender CM. {Introduction to PT-symmetric quantum theory}. \emph{Contemp. Phys.}
  46(4), 277,  2005

\bibitem{bender13b}
Bender CM. {Making sense of non-Hermitian Hamiltonians}. \emph{Reports Prog.
  Phys.} 70(6), 947,  2007

\bibitem{Mostafazadeh2012}
Mostafazadeh A. {Invisibility and PT symmetry}. \emph{Phys. Rev. A - At. Mol.
  Opt. Phys.} 87(1), 012103,  2013

\bibitem{Ge2012}
Ge~L, Chong YD, Stone AD. {Conservation relations and anisotropic transmission
  resonances in one-dimensional PT-symmetric photonic heterostructures}.
  \emph{Phys. Rev. A - At. Mol. Opt. Phys.} 85(2), 1,  2012

\bibitem{chong2011}
Chong YD, Ge~L, Stone AD. {PT-symmetry breaking and laser-absorber modes in
  optical scattering systems}. \emph{Phys. Rev. Lett.} 106(9), 093902,  2011

\bibitem{longhi2010a}
Longhi S. {PT-symmetric laser absorber}. \emph{Phys. Rev. A} 82(3), 031801,
  2010

\bibitem{ISI:000266685400002}
Mostafazadeh A. {Spectral singularities of complex scattering potentials and
  infinite reflection and transmission coefficients at real energies}.
  \emph{Phys. Rev. Lett.} 102(22), 220402,  2009

\bibitem{Benson1991}
Benson FA, Benson TM, \emph{{Fields, Waves and Transmission Lines}} (Springer,
  Amsterdam, 1991)

\bibitem{Ramo199}
Ramo S, Whinnery JR, Duzer TV, \emph{{Fields and Waves in Communication
  Electronics}} 3rd edn. (John Wiley, 1999)

\bibitem{Pozar2011}
Pozar DM, \emph{{Microwave Engineering}} 4th edn. (John Wiley, New York, NY,
  2011)

\bibitem{Haus1983}
Haus HA, \emph{{Waves and Fields in Optoelectronics}} (Prentice-Hall, New
  Jersey, 1983)

\bibitem{Collin1991}
Collin RE, \emph{{Field Theory of Guided Waves}} 2nd edn. (IEEE Press, New
  York, NY, 1991)

\bibitem{Jalas2013}
Jalas D, Petrov A, Eich M, Freude W, Fan S, Yu~Z, Baets R, Popovi{\'{c}} M,
  Melloni A, Joannopoulos JD, Vanwolleghem M, Doerr CR, Renner H. {What is -
  and what is not - an optical isolator}. \emph{Nat. Photonics} 7(8), 579,
  2013

\bibitem{Iizuka2002}
Iizuka K, \emph{{Elements of Photonics, Vol II}} (John Wiley, New York, NY,
  2002)

\bibitem{Saleh2007}
Saleh BEA, Teich MC, \emph{{Fundamentals of Photonics}} 2nd edn. (John Wiley,
  New York, NY, 2007)

\bibitem{Liu2005}
Liu JM, \emph{{Photonic Devices}} (Cambridge University Press, Cambridge, 2005)

\bibitem{Hagness1996}
Hagness SC, Joseph RM, Taflove A. {Subpicosecond electrodynamics of distributed
  Bragg reflector microlasers: Results from finite difference time domain
  simulations}. \emph{Radio Sci.} 31(4), 931,  1996

\bibitem{Siegman1986}
Siegman AE, \emph{{Lasers}} (University Science Book, Palo Alto, CA, 1986)

\bibitem{mich2015}
Robertson M.
\newblock Private communication, October 2015

\bibitem{Landau1984}
Landau LD, Bell JS, Kearsley MJ, Pitaevskii LP, Lifshitz EM, Sykes JB,
  \emph{{Electrodynamics of Continuous Media}} 2nd edn. (Elsevier, London,
  England, 1984)

\bibitem{Zyablovsky2014}
Zyablovsky AA, Vinogradov AP, Dorofeenko AV, Pukhov AA, Lisyansky AA.
  {Causality and phase transitions in PT-symmetric optical systems}.
  \emph{Phys. Rev. A} 89(3), 033808,  2014

\bibitem{Hoefer1985}
Hoefer W. {The transmission-line matrix method--theory and applications}.
  \emph{IEEE Trans. Microw. Theory Tech.} 33(10), 882,  1985

\bibitem{Christopoulos1995}
Christopoulos C, \emph{{The Transmission-Line Modeling Method TLM}} (IEEE
  Press, Piscataway, 1995)

\bibitem{Paul1998}
Paul J, {Modelling of general electromagnetic material properties in TLM}.
\newblock Ph.D. thesis University of Nottingham 1998

\bibitem{Paul2002}
Paul J, Christopoulos C, Thomas D. {Generalized material models in TLM - part
  III: Materials with nonlinear properties}. \emph{IEEE Trans. Antennas
  Propag.} 50(7), 997,  2002

\bibitem{Paul1999a}
Paul J, Christopoulos C, Thomas D. {Generalized material models in TLM - part
  I: Materials with frequency-dependent properties}. \emph{IEEE Trans. Antennas
  Propag.} 47(10), 1528,  1999

\bibitem{Janyani2004a}
Janyani V, Vukovic A, Paul J. {The development of TLM models for nonlinear
  optics}. \emph{Microw. Rev.} 10(1), 35,  2004

\bibitem{Janyani2005a}
Janyani V, Vukovic A, Paul JD, Sewell P, Benson TM. {Time domain simulation in
  photonics: A comparison of nonlinear dispersive polarisation models}.
  \emph{Opt. Quantum Electron.} 37(1-3), 3,  2005

\bibitem{Meng2013}
Meng X, Sewell P, Vukovic A, Dantanarayana HG, Benson TM. {Efficient broadband
  simulations for thin optical structures}. \emph{Optical and Quantum
  Electronics} 45(4), 343,  2013

\bibitem{Meng2015}
Meng X, Sewell P, Phang S, Vukovic A, Benson TM. {Modeling Curved Carbon Fiber
  Composite (CFC) Structures in the Transmission-Line Modeling (TLM) Method}.
  \emph{IEEE Transactions on Electromagnetic Compatibility} 57(3), 384,  2015

\bibitem{Joseph1997}
Joseph R, Taflove A. {FDTD Maxwell's equations models for nonlinear
  electrodynamics and optics}. \emph{IEEE Trans. Antennas Propag.} 45(3), 364,
  1997

\bibitem{FatkhullaKh.Abdullaev2005}
{Fatkhulla KhA} , Konotop VV, \emph{{Nonlinear Waves: Classical and Quantum
  Aspects}} \emph{NATO Science Series II: Mathematics, Physics and Chemistry},
  vol. 153 (Springer Netherlands, Dordrecht, 2005).
\newblock \doi{10.1007/1-4020-2190-9}

\bibitem{Conti2004}
Conti C, {Di Falco} A, Assanto G. {Optical parametric oscillations in isotropic
  photonic crystals}. \emph{Opt. Express} 12(5), 823,  2004

\bibitem{Janyani2004}
Janyani V, Paul J, Vukovic A, Benson T, Sewell P. {TLM modelling of nonlinear
  optical effects in fibre Bragg gratings}. \emph{IEE Proc. - Optoelectron.}
  151(4), 185,  2004

\bibitem{Kazantseva2005}
Kazantseva EV, Maimistov AI, Caputo JG. {Reduced Maxwell-Duffing description of
  extremely short pulses in nonresonant media}. \emph{Phys. Rev. E} 71(5),
  056622,  2005

\bibitem{Press2002}
Press WH, Teukolsky SA, Vetterling WT, Flannery BP, \emph{{Numerical Recipes in
  C++: The Art of Scientific Computing}} 2nd edn. (Cambridge University Press,
  Cambridge, 2002)

\bibitem{Suryanto2003}
Suryanto A, Groesen Evan, Hammer M, Hoekstra HJWM. {A finite element scheme to
  study the nonlinear optical response of a finite grating without and with
  defect}. \emph{Opt. Quantum Electron.} 35(1997), 313,  2003

\bibitem{Brzozowski2000}
Brzozowski L, Sargent E. {Optical signal processing using nonlinear distributed
  feedback structures}. \emph{IEEE J. Quantum Electron.} 36(5), 550,  2000

\end{thebibliography}

\printindex
\end{document}